\documentclass[reprint,superscriptaddress,amsmath,amssymb,aps]{revtex4-1}

\usepackage{graphicx}
\begin{document}

\title{Experimentally-realizable $\mathcal{PT}$ phase transitions in reflectionless quantum scattering}

\author{Micheline B.~Soley}

\affiliation{Department of Chemistry, University of Wisconsin-Madison, 1101 University Ave., Madison, WI 53706}

\affiliation{Department of Physics, University of Wisconsin-Madison, 1150 University Ave., Madison, WI 53706}

\affiliation{Yale Quantum Institute, Yale University, PO Box 208334, New Haven,
CT 06520, USA}

\affiliation{Department of Chemistry, Yale University, 225 Prospect St., New Haven,
CT 06520, USA}

\author{Carl M.~Bender}

\affiliation{Department of Physics, Washington University, St. Louis, MO 63130, USA}

\author{A.~Douglas Stone}

\affiliation{Yale Quantum Institute, Yale University, PO Box 208334, New Haven,
CT 06520, USA}

\affiliation{Department of Applied Physics, Yale University, New Haven, CT 06520, USA}

\begin{abstract}

A class of above-barrier quantum-scattering problems is shown to provide an experimentally-accessible platform for studying $\mathcal{PT}$-symmetric Schr{\"o}dinger equations that exhibit spontaneous $\mathcal{PT}$ symmetry breaking despite having purely real potentials. These potentials are one-dimensional, inverted, and unstable and have the form $V(x) = - \lvert x\rvert^p$ ($p>0$), terminated at a finite length or energy to a constant value as $x\to \pm\infty$.
The signature of {\em unbroken} $\mathcal{PT}$ symmetry is the existence of reflectionless propagating states at discrete real energies up to arbitrarily high energy. 
In the $\mathcal{PT}$-{\em broken} phase, 
there are no such solutions.  In addition, there exists an intermediate mixed phase, 
where reflectionless states exist at low energy but disappear at a fixed finite energy, independent of termination length. In the mixed phase exceptional points (EPs) occur at specific $p$ and energy values, with a quartic dip in the reflectivity in contrast to the quadratic behavior away from
EPs.  $\mathcal{PT}$-symmetry-breaking phenomena have not been previously predicted in a quantum system with a real potential and no reservoir coupling. The effects predicted here are measurable in standard cold-atom experiments with programmable optical traps. The physical origin of the symmetry-breaking transition is elucidated using a WKB force analysis that identifies the spatial location of the above-barrier scattering.

\end{abstract}

\maketitle

Above-barrier reflection is a fundamental quantum effect in which particles with sufficient energy to surmount a potential-energy barrier are nonetheless reflected backwards.  It can be seen as complementary to the phenomenon of quantum tunneling, in which a particle that lacks sufficient energy to overcome a barrier is nonetheless able to pass through the barrier with some probability. Both effects arise from the fundamental wave character of the quantum wavefunction, which cannot change abruptly with small changes in the potential. For a typical potential barrier with maximum energy $V_0$, a quantum particle will reflect with high probability when its energy $E$ is slightly higher than $V_0$, and at higher energies the reflection probability decreases rapidly on average.  However, it has been known for quite some time that for certain shapes of potential this decrease is monotonic, so that unit transmission is only achieved at infinite energy.  For other shapes the decrease is nonmonotonic, and there exist discrete energies near $V_0$ at which perfect above-barrier transmission is possible.
In this work we elucidate this difference for a large class of potential shapes by relating it to the existence of broken or unbroken parity-time ($\mathcal{PT}$) symmetry in the corresponding Schr{\"o}dinger equation. Experiments testing the predictions of our approach are feasible using cold atomic beams and current atomic-trap technology.

Restricting our discussion to one-dimensional potentials $V(x)$, $\mathcal{PT}$ symmetry refers to Schr{\"o}dinger equations and boundary conditions that map into themselves under combined $x \to -x$ and complex conjugation.  This $\mathcal{PT}$-symmetry condition is less restrictive than the Hermiticity condition imposed in conventional quantum mechanics and allows for complex potentials with imaginary parts antisymmetric around the origin. Essentially all research on this topic has focused on Schr{\"o}dinger equations with complex potentials.  Here we treat Schr{\"o}dinger equations with {\em real} potentials that have only $\mathcal{PT}$ symmetry and are not Hermitian due to the reflectionless boundary conditions.

Because investigations of $\mathcal{PT}$ symmetry in quantum mechanics have typically involved complex potentials, which involve imaginary terms that cannot be implemented in closed quantum systems, there have been few experimental demonstrations of phenomena related to $\mathcal{PT}$ symmetry in quantum mechanics.  Those few have been restricted to open systems coupled to reservoirs that are treated statistically and introduce phenomenological terms into the Schr{\"o}dinger equation \cite{Zhao.2010.042903,Bittner.2012.024101,Zheng.2013.20120053}.  
In contrast, there has been an intensive experimental study of $\mathcal{PT}$ symmetry and its breaking in systems described by
classical wave equations, where imaginary terms representing loss and gain are introduced into the equations to represent coupling to reservoirs, but these tend to be more controllable and relatively easy to fabricate and measure.  Examples span classical electromagnetism \cite{Guo.2009.093902,Ruter.2010.192,Feng.2011.729,Regensburger.2012.167,Xiao.2021.230402,Peng.2014.394}, acoustics \cite{Shi.2016.1,Auregan.2017.174301}, electronics \cite{Chtchelkatchev.2012.150405,Schindler.2011.040101,Bender.2013.234101,Cao.2022.1}, and mechanical systems \cite{Bender.2013.173}. In several cases these classical $\mathcal{PT}$-symmetric systems have shown potential utility for applications in laser technology \cite{Feng.2014.972,Hodaei.2014.975}, sensing \cite{chen2016p,liu2016metrology}, and wireless power transfer \cite{Assawaworrarit.2017.387,Assawaworrarit.2020.273}.

Our work builds off the work in Refs.~\cite{Bender.1998.5243,Bender.2007.947}, which studied $\mathcal{PT}$-symmetry phenomena in a class of quantum systems with complex potentials of the form $V(x)={x}^2(\text{i}{x})^\epsilon$ and $V(x)={x}^4(\text{i}{x})^\epsilon$ ($\epsilon$ real), all of which satisfy the $\mathcal{PT}$-symmetry condition. Importantly, for even integer values of $\epsilon$, this set includes purely real ``upside-down" $\mathcal{PT}$-symmetric potentials (specifically $V(x)=-x^2, -x^4,-x^6, -x^8,\ldots$). It was shown that for these real potentials with even integer power $p \geq 4$ there exist discrete weakly-bound states for real energies $E_i >0$, but for $p=2$ there are no real-energy solutions, only complex-energy ones. It was realized that real energies correspond to a kind of reflectionless-scattering state \cite{ahmed2005reflectionless}, although due to the unbounded-below nature of the potentials, the particles would be moving arbitrarily fast as they approached $\pm \infty$, causing the wavefunctions to oscillate infinitely fast and invalidating the assumptions of standard scattering theory. This paper proposes a clear way to probe this quantum physics in an experimentally-realizable setup.

Here we show that the reflectionless states of a continuous class of {\em truncated} upside-down real potentials show all the characteristics of $\mathcal{PT}$-symmetric systems; specifically the presence of unbroken, mixed, and broken phases and spontaneous symmetry-breaking transitions at exceptional points, despite the absence of any imaginary terms in the potential. We quantitatively confirm the presence of weakly-bound states above such potentials with energies given precisely by those predicted for the integer cases considered in Ref.~\cite{Bender.2018.052118}. This constitutes the first theoretical prediction of such $\mathcal{PT}$-symmetry phenomena in experimentally-realizable {\em quantum} scattering systems governed by the Schr{\"o}dinger equation.

Here we consider potentials of the form $V(x)=-\lvert x\rvert ^p$ for $p\in\mathbb{R}$, truncated to constant energy outside of the domain $-L\le x\le L$ (see Fig.~\ref{fig:Potentials}); we consider energy truncation instead of length truncation in the Appendix.  These potentials are real and parity symmetric and are nonanalytic at the origin for noneven $p$ values. To yield a continuously differentiable potential at the truncation point, we introduce a smoothed potential $V\left(x,w\right)$, where the real parameter $w$ determines the sharpness of the truncation (see Methods).  
 $V\left(x,w\right)$ reproduces the original upside-down $\mathcal{PT}$-symmetric potentials in a finite region close to $x=0$, where the eigenmode amplitude is expected to be greatest. Motivated by a different truncation used in ref.~\cite{Bender.2018.052118}, our approach is qualitatively different, as it generalizes the original potentials to noneven-integer powers and truncates them to yield a standard quantum-scattering geometry. The nonanalytic behavior distinguishes this class of $\mathcal{P}$- and $\mathcal{T}$-symmetric real potentials from the aforementioned $\mathcal{PT}$-symmetric potentials $V(x)=x^2(\text{i}x)^\epsilon$ and $V(x)=x^4(\text{i}x)^\epsilon$, which can be continued in the complex plane, whereas $V(x)=-\lvert x \rvert^p$ cannot. Note that these noneven-integer potentials are analytic in $p$, but not in $x$; hence, the eigenvalues are analytic functions of $p$, but the corresponding Schr{\"o}dinger equation is not analytic in $x$. 

\begin{figure}
\centering
\includegraphics[width=1\columnwidth]{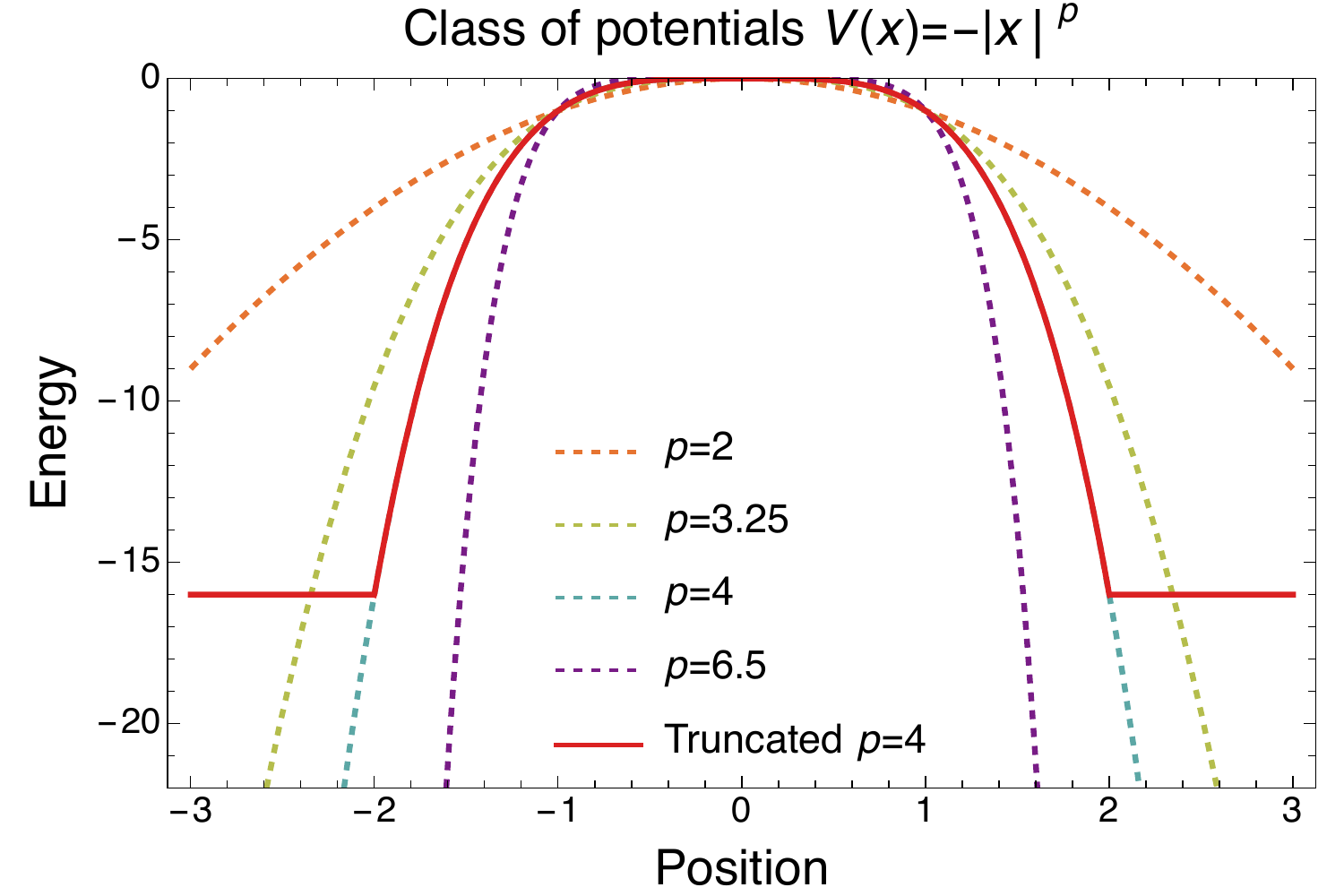}
\caption{Upside-down, unbounded-below, $\mathcal{PT}$-symmetric potential $-x^4$ (dashed blue line) is shown as a member of the larger class of purely real potentials $V(x)=-\lvert x\rvert ^p$ for $p\in\mathbb{R}$ (dashed lines indicate instances $p=2$ [orange], $3.25$ [yellow], and $6.5$ [purple]). The solid red line illustrates our truncation of these potentials for $p=4$ to length $L=2$ and smoothing parameter $w=1000$. We study below the resulting above-barrier scattering phenomena and their relation to the corresponding unbounded, infinite system. \label{fig:Potentials}}
\end{figure}  

We first verify that the reflectionless states of these truncated $V(x)=-V_0\lvert x\rvert ^p$ potentials obey $\mathcal{PT}$ symmetry. We introduce $V_0$ to clarify the choice of units. Since the potential is real, it is both $\mathcal{P}$- and $\mathcal{T}$-symmetric, and the problem superficially appears to be Hermitian; however, we shall see that the symmetry is reduced by the reflectionless boundary conditions. We start with the one-dimensional Schr{\"o}dinger equation
\begin{align}
0&=-\frac{\hbar^2}{2m}\phi^{\prime\prime}(x)+\left[V(x)-E\right]\phi(x),\nonumber
\end{align}
 where $m$ is the mass of the relevant quantum particle. We introduce the length scale $x_0 \equiv  (E/V_0)^{1/p}$, where for values of $p$ for which the infinite potential has bound states we choose $E=E_0$, the ground-state energy, so that $V(x)= -E_0 \lvert x\rvert^{p} $. (In the case of no bound states $E$ can be chosen arbitrarily.) Henceforth, $x$ denotes the position in units of $x_0$ and we take energy units with $E_0=1$ to recover $V(x)=-\lvert x\rvert ^p$.
Assuming that the potential and kinetic energy are of the same order and hence of order $E_0$, we find that $x_0 \sim (h^2 /m V_0)^{{\frac{1}{p+2}}}, E_0 \sim V_0 x_0^p.$
Now we look for a solution with a right-moving wave only, which satisfies the boundary conditions
\begin{equation}
    \phi^\prime(-L)\sim\text{i}k\phi(-L), \quad \phi^\prime(L)\sim\text{i}k\phi(L)\quad(w\to0),\nonumber
\end{equation}
where $k \equiv \sqrt{2mE/\hbar}$.
These conditions map into left-moving waves under either $\mathcal{P}$ or $\mathcal{T}$ separately, but map into themselves under the product $\mathcal{PT}$. The well-known implication is that any solutions must either have a real energy or must occur in complex-conjugate pairs. If the former holds for all solutions, the $\mathcal{PT}$ symmetry is said to be {\em unbroken}; if the latter holds, it is {\em broken}.  If both types of solutions exist, the spectrum is mixed. 

For the infinite-length and depth potentials studied in Ref.~\cite{Bender.2007.947}, the real solutions are peaked at the origin and decay weakly [$\phi(x) \sim 1/x$ as $x \to \pm\infty$ for the case $p=4$]. This contrasts strongly with the standard exponentially-decaying states of real attractive potentials.  For the finite potentials (after truncation) and the scattering geometry, the reflectionless above-barrier states propagate as plane waves at infinity and are not square integrable.
A recent general theory of reflectionless scattering modes (RSMs) \cite{Sweeney.2020.063511,Stone.2020.343} predicts an infinite set of discrete solutions for generic potentials with $\mathcal{PT}$ symmetry that also occur at real energies or in complex-conjugate pairs.  
We conjecture that these propagating reflectionless solutions will mimic closely the weakly-bound states of the unbounded potentials in the vicinity of the origin and will occur at similar energies. 

The hypothesis is verified by applying the RSM theory to quantum mechanics to determine accurately the reflectionless scattering modes. Until now, this theory has been applied to electromagnetic/optical systems \cite{Sweeney.2020.063511,Stone.2020.343}, but the method is general and can be applied to quantum-scattering systems in any number of dimensions for scattering of both free-space and guided waves. We derive a specifically quantum formalism for the RSM theory in Appendix~\ref{sec:RSMforQuantum}. An attractive  feature of the theory is that it calculates directly the discrete complex spectrum of reflectionless energies (referred to as {\em R-zeros}). There is no need to solve the scattering problem and search for zero reflection. The R-zero spectrum is similar to the more familiar spectrum of complex energy resonances, which satisfy purely outgoing boundary conditions; but the R-zero energies are distinct from the resonances. A sketch of the RSM/R-zero theory for general scattering geometries is given in the Methods.  
 
The general theory simplifies for the current case of a one-dimensional (two-channel) geometry, and the reflectionless energies can be found by a simple modification of the method of perfectly-matched layers, complex scaling \cite{moiseyev1998quantum}, or complex absorbing potentials \cite{muga2004complex},
which we use here.  Real-energy solutions correspond to steady-state harmonic scattering and thus imply a zero of the reflection coefficient at the correct input energy. Complex solutions do not give zero reflection under uniform harmonic excitation but, if the R-zero is isolated and near the real axis, there is a narrow dip in the reflection below the background near the real part of its energy.  Hence we can study these R-zero spectra and the corresponding eigenfunctions for the truncated potentials introduced and compare their properties to the infinite unbounded potentials studied previously.  We find striking agreement between the two systems, bringing the novel predictions of the previous theory into experimental reach.

\begin{figure*}
\centering
\includegraphics[height=0.25\textheight]{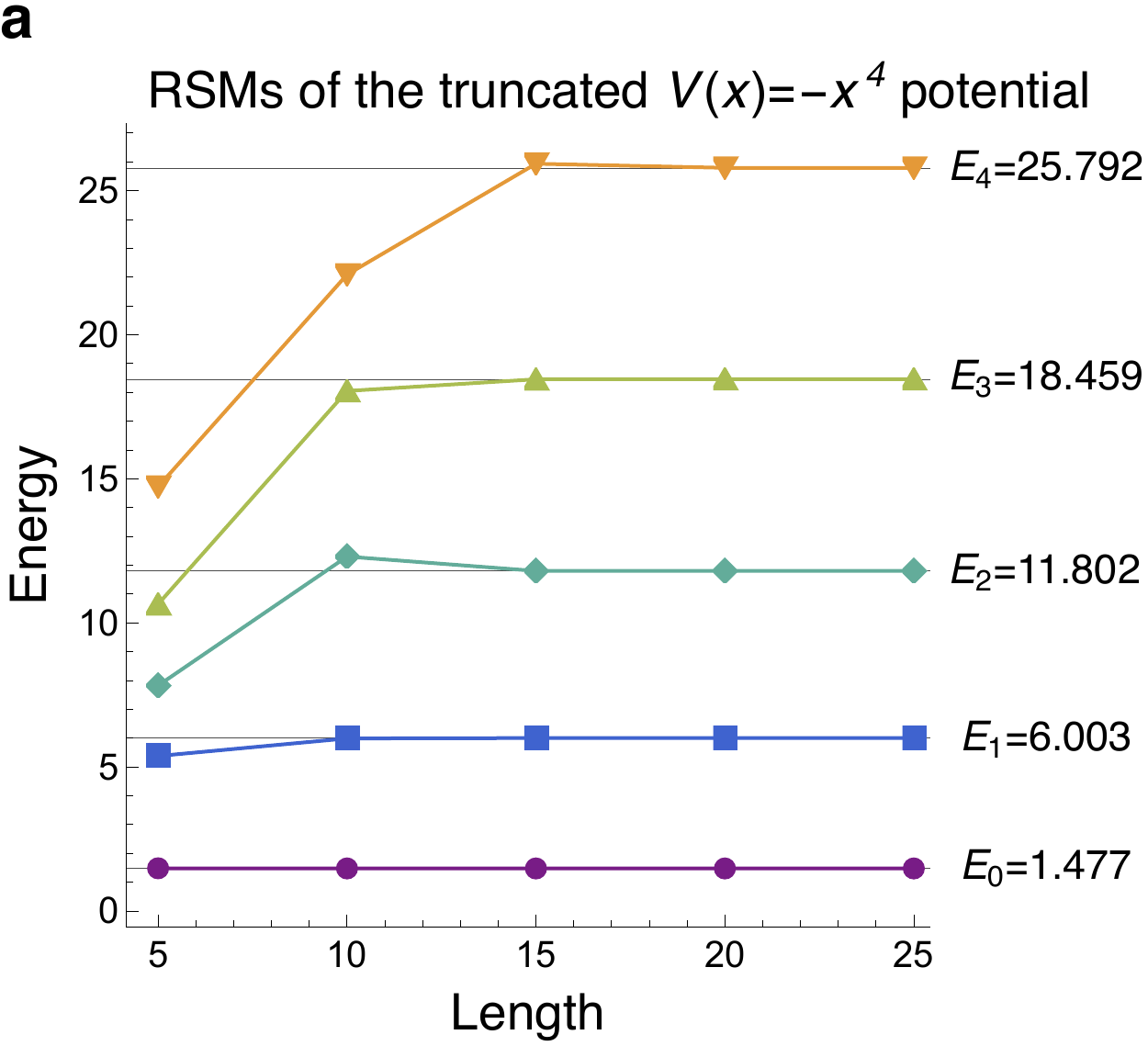}\,\includegraphics[height=0.25\textheight]{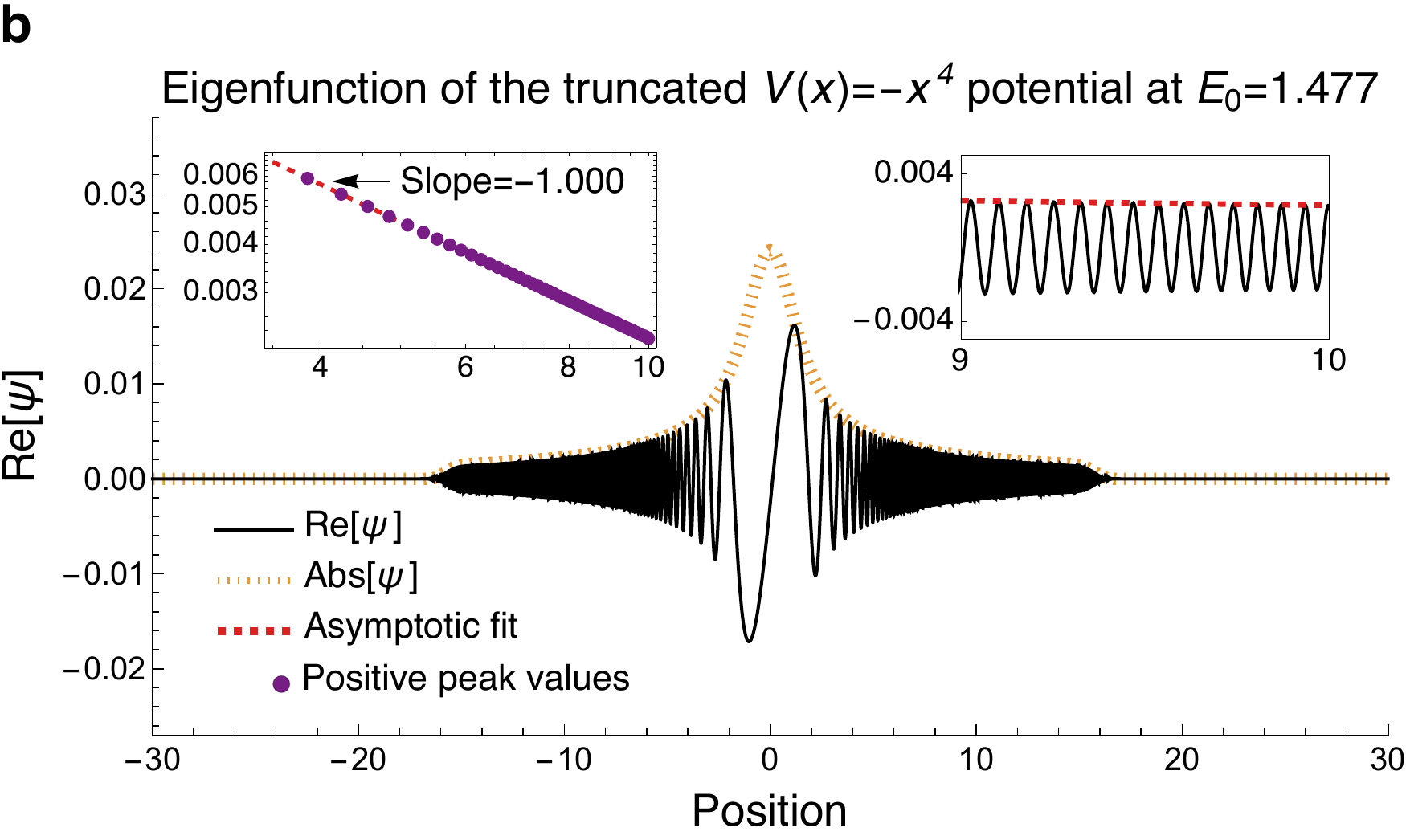}
\caption{(a) Reflectionless scattering modes (RSMs) of the truncated $V(x)=-x^4$ potential are found to have real energies that converge to the exact analytic weakly-bound-state energies $E_i$ of the corresponding infinite-length potential (gray horizontal lines, $i=0-4$ shown) for sufficient length $L$. (b) The scattering wavefunction corresponding to RSMs of the truncated $V(x)=-x^4$ potential (real part, black solid line; absolute value, dashed orange line) for $E=1.477$. Note that the modulus of the wavefunction exhibits parity symmetry, and the real part has an accelerating oscillatory behavior away from the origin. The envelope obeys an asymptotic power law in precise agreement with predictions for the weakly bound states of the infinite-length $V(x)=-x^4$ potential, fit by $\psi_\text{env}=0.0228x^{-1.000}$   (dashed red line). \label{fig:RSMTruncatedResults}}
\end{figure*}

As shown in Fig.~\ref{fig:RSMTruncatedResults}(a) and Appendix~\ref{sec:BoundedEnergy}, the low-energy RSM energies for $p=4$  are indeed real and agree with the bound-state energies found for the infinite system with 7-8 digits of accuracy for sufficiently large $L$. Even for the shortest truncation length $L$ considered, the ground-state energy is accurately found, and higher eigenenergies converge to known infinite system values with increasing length $L$. Once the incident energies are known, these results are easily tested and confirmed by quantum-scattering calculations, as detailed in Appendix~\ref{sec:QuantumScatteringTruncation}. Similar convergence is found with results obtained by energy truncation at $\lvert V_\text{max}\rvert $ (Appendix~\ref{sec:BoundedEnergy}); see also more general scattering results in Fig.~3(b).

Not only are the energies accurately predicted  \cite{Bender.2018.052118}, but also the eigenfunctions in the interaction region perfectly mimic the behavior predicted for the infinite system. As shown in Fig.~\ref{fig:RSMTruncatedResults}(b), the eigenfunctions are symmetric around the origin and exhibit the predicted $1/x$ decay with three-digit accuracy. Thus, the intriguing weakly-bound states of the infinite potential are directly observable in the truncated system.  Note the rapid increase in the spatial oscillation frequency of the eigenfunction as the particle accelerates toward the asymptotic region, where the force vanishes.

\begin{figure*}
\centering
\includegraphics[width=0.44\textwidth]{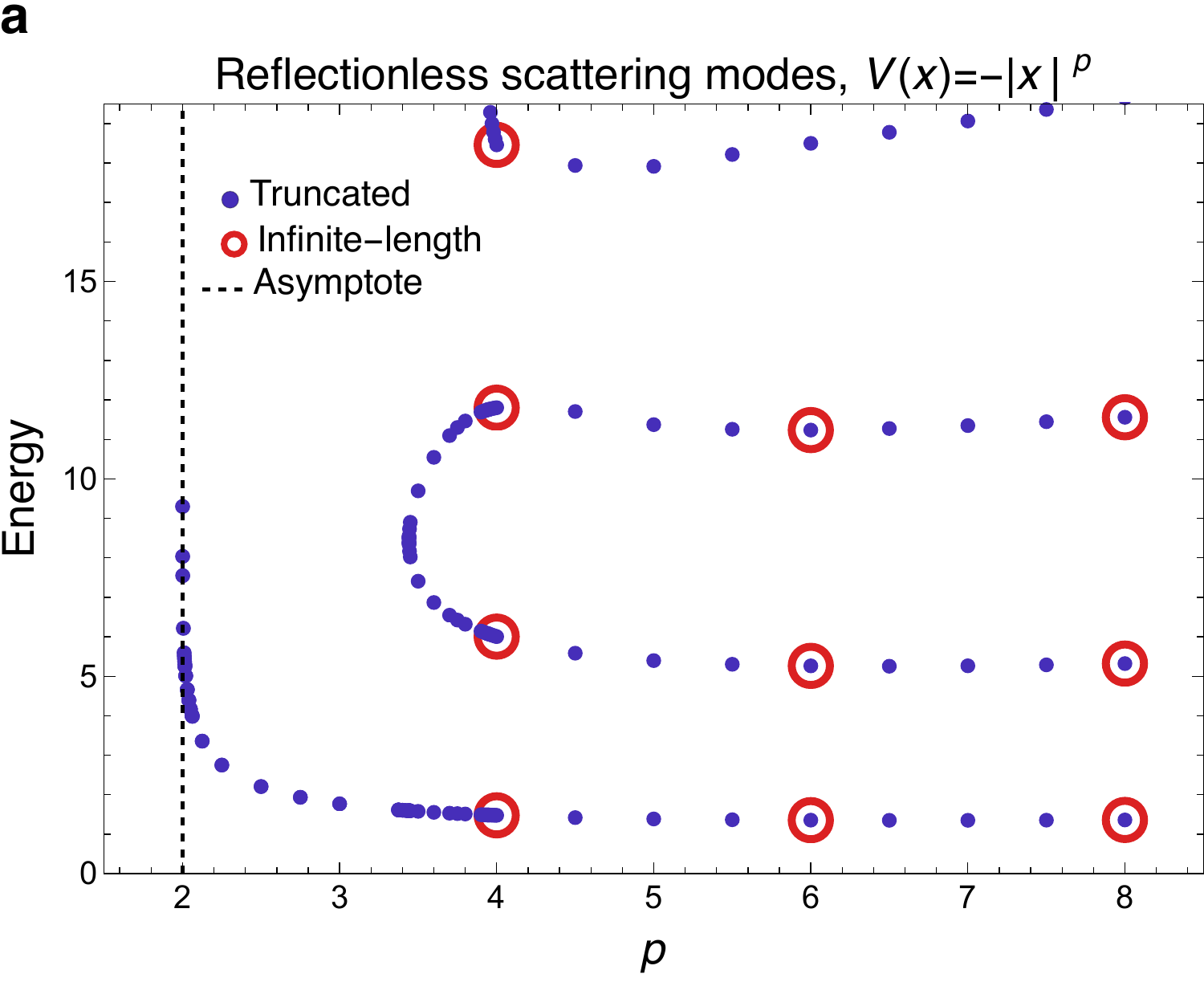}\qquad\includegraphics[width=0.44\textwidth]{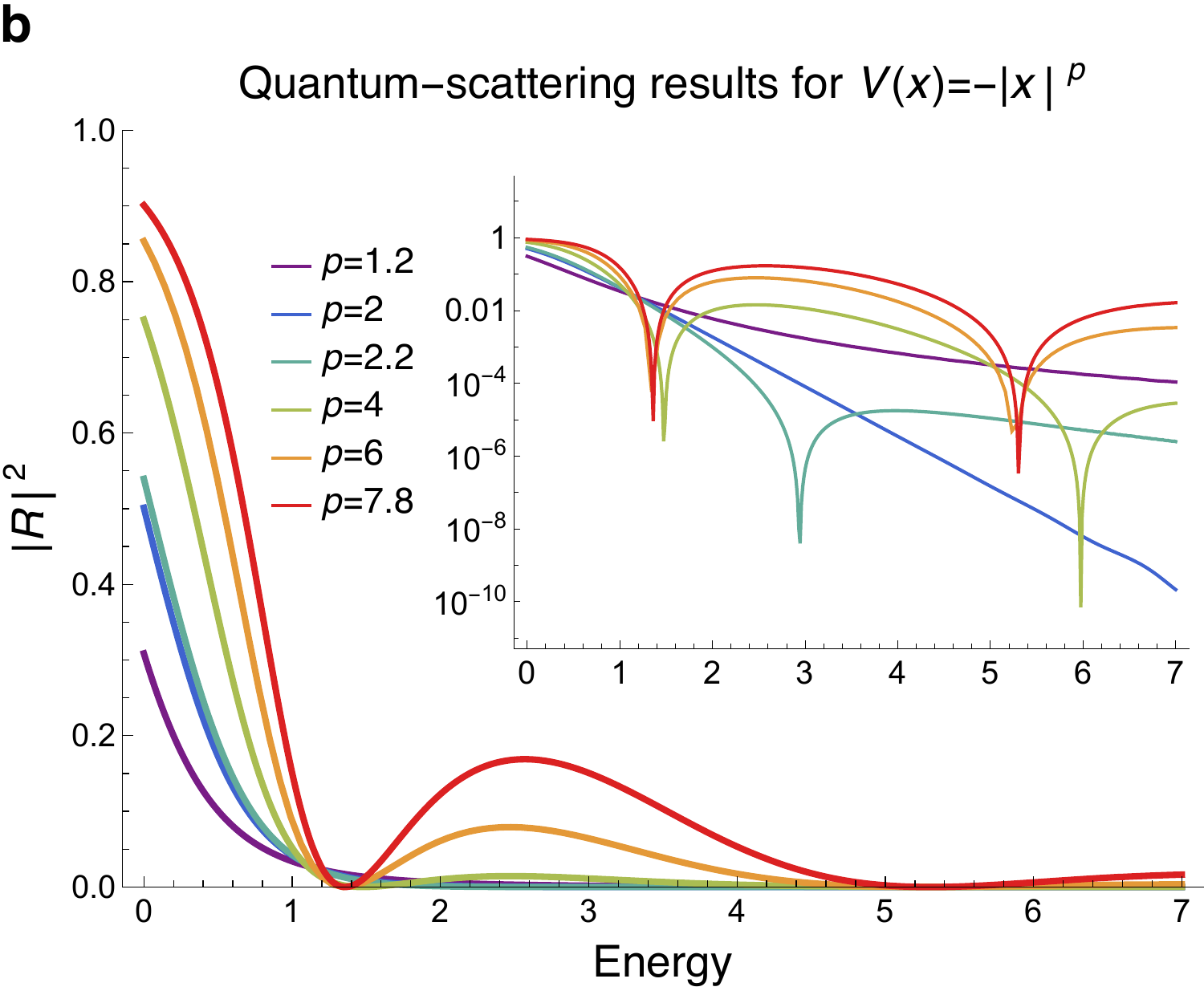}
\caption{(a) RSM spectrum of the class of truncated $V(x)=-\lvert x\rvert ^p$ potentials and (b) quantum-scattering results for same. In (a) results indicate no RSMs exist as $p \to 2$ from above. The single remaining RSM for $p<3.44$ is pushed asymptotically to infinity as $p \to2$, in agreement with the result for complex potentials \cite{Bender.2007.947}. For $p \leq 2$ we find no RSMs, corresponding to a fully-broken $\mathcal{PT}$ phase.  There is a mixed phase region for $2 < p \leq4$;  and many RSMs for $p\ge 4$, with the number increasing with the increasing truncation length, indicating an unbroken $\mathcal{PT}$ phase.
The lower energy RSMs agree closely with the known energies of the weakly-bound states of the infinite-length potentials for even $p$ (red circles). (b) Quantum-scattering results show monotonic decay of the reflectance with energy for $p \leq 2$  and nonmonotonic behavior for $p>2$ with deep dips at the energies of the RSMs. 
For $p>2$ there is a second reflection peak after the first dip, which is substantial ($\sim 0.10)$ for $p=6,7.8$. \label{fig:RSMTransitions}}
\end{figure*}

RSM theory also allows us to search for a $\mathcal{PT}$ quantum phase transition within the class of potentials $V(x)=-\lvert x\rvert ^p$ as $p$ is varied, a phenomenon suggested by earlier work on complex $\mathcal{PT}$ potentials cited above \cite{Bender.1998.5243,Bender.2007.947}.
In Fig.~\ref{fig:RSMTransitions}(a) we plot the low-lying real-energy eigenvalues of this class of potential as $p$ is varied between $2 \leq p \leq 8$.  The $\mathcal{PT}$ symmetry implies that as $p$ is varied, eigenvalues cannot disappear, but can meet at certain parameter values and then generically move into the complex plane as conjugate pairs.  We find that for $p \geq 4$ there are discrete real energies up to a high energy, above which the effect of our truncation becomes visible.  This indicates that the behavior found for the infinite system at the discrete integers $p=4,6,8$ (infinite number of real-energy bound states) is generic and continuous in $p$. 

In the interval $2<p<4$ we see clearly a series of eigenvalue intersections, which happen at higher $p$ for higher energies. These are {\em exceptional points} (EPs), where two distinct eigenfunctions coalesce before the energies become complex (we only plot real energies here).  
In the interval  $2<p<4$ there are only a finite number of real reflectionless energies, corresponding to partially-broken $\mathcal{PT}$ symmetry, similar but distinct from the behavior of the real eigenenergies of the complex potentials \cite{Bender.2007.947}, $V(x)=x^2(\text{i}x)^\epsilon$. 
Finally, for $p\le 2$, there are {\em no} real reflectionless energies, and the reflection coefficient simply decays monotonically with increasing energy. 
 This monotonically-decreasing, nonresonant behavior of scattering above the inverted harmonic potential has been known since the early days of quantum mechanics \cite{kemble1935contribution}. Now we can appreciate that the behavior changes qualitatively  for any sharper polynomial barrier, and that the regime $p\leq 2$ has fully-broken $\mathcal{PT}$ symmetry, with no real-energy reflectionless states.  
As for $p=4$, our results for $p=2$ agree very well with the known results for the infinite potentials for sufficiently long truncation length $L$ (see Appendix~\ref{sec:QuantumScatteringTruncation}).

Quantum-scattering results shown in Fig.~\ref{fig:RSMTransitions}(b) dramatically confirm the identification of a distinct transition between the fully-broken $\mathcal{PT}$ phase ($p\le 2$) and the partially-broken phase ($2<p\leq 4$). Up to $p=2$, we see a slower than exponential, but monotonic, decay of the reflection coefficient as a function of energy, without any deep dips.  At $p=2$, the asymptotic decay is exponential, and above it the decay is nonmonotonic with deep dips at the predicted reflection zeros (resolved only to a finite depth due to the finite accuracy of the numerics).  Note also that the peak reflection for $p=6$, \emph{after} the first zero, has a reflection coefficient of $\sim 0.10$, which should be easily measurable in experiments.  

\begin{figure*}
\centering
\includegraphics[width=0.44\textwidth]{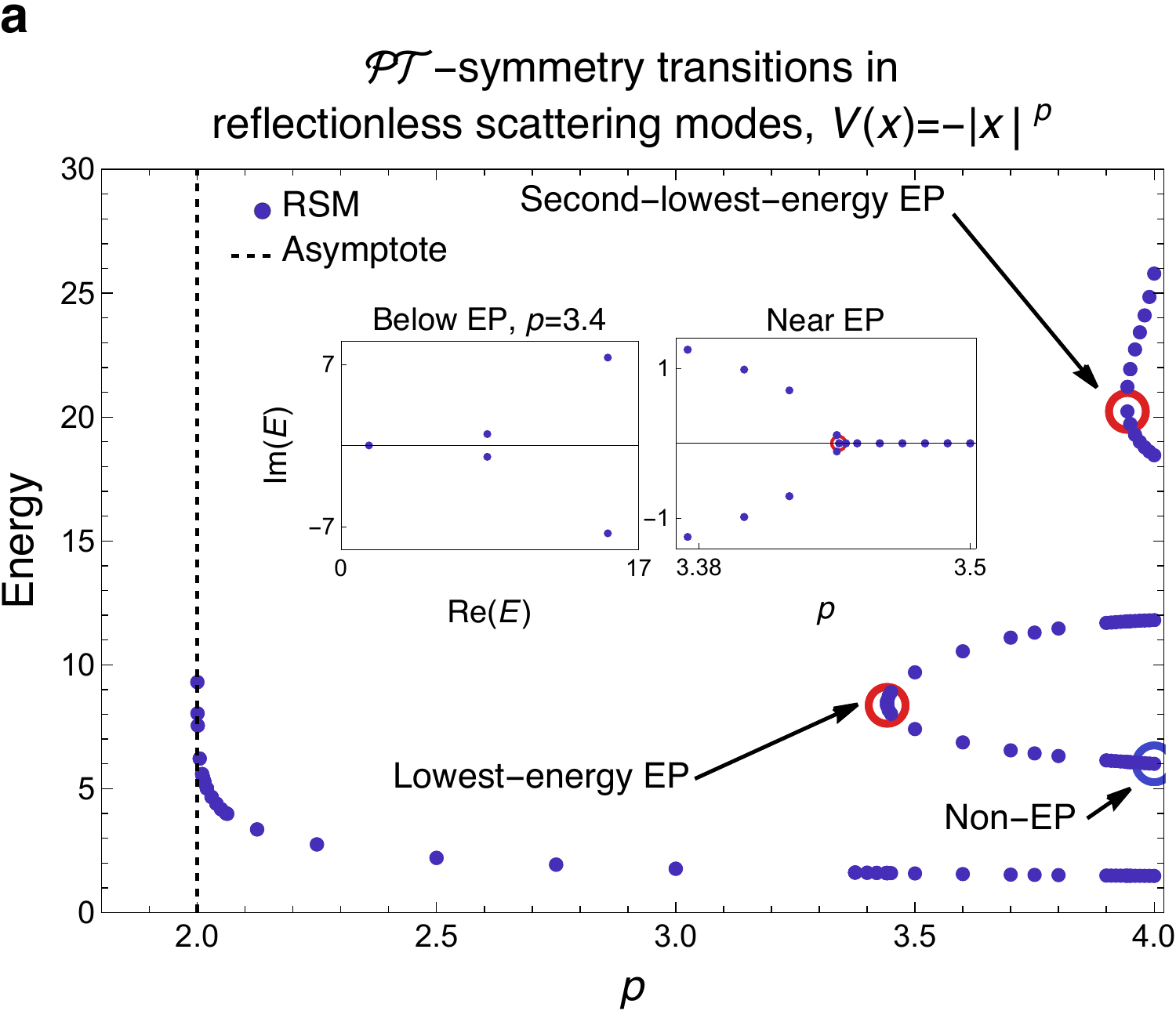}\qquad\includegraphics[width=0.44\textwidth]{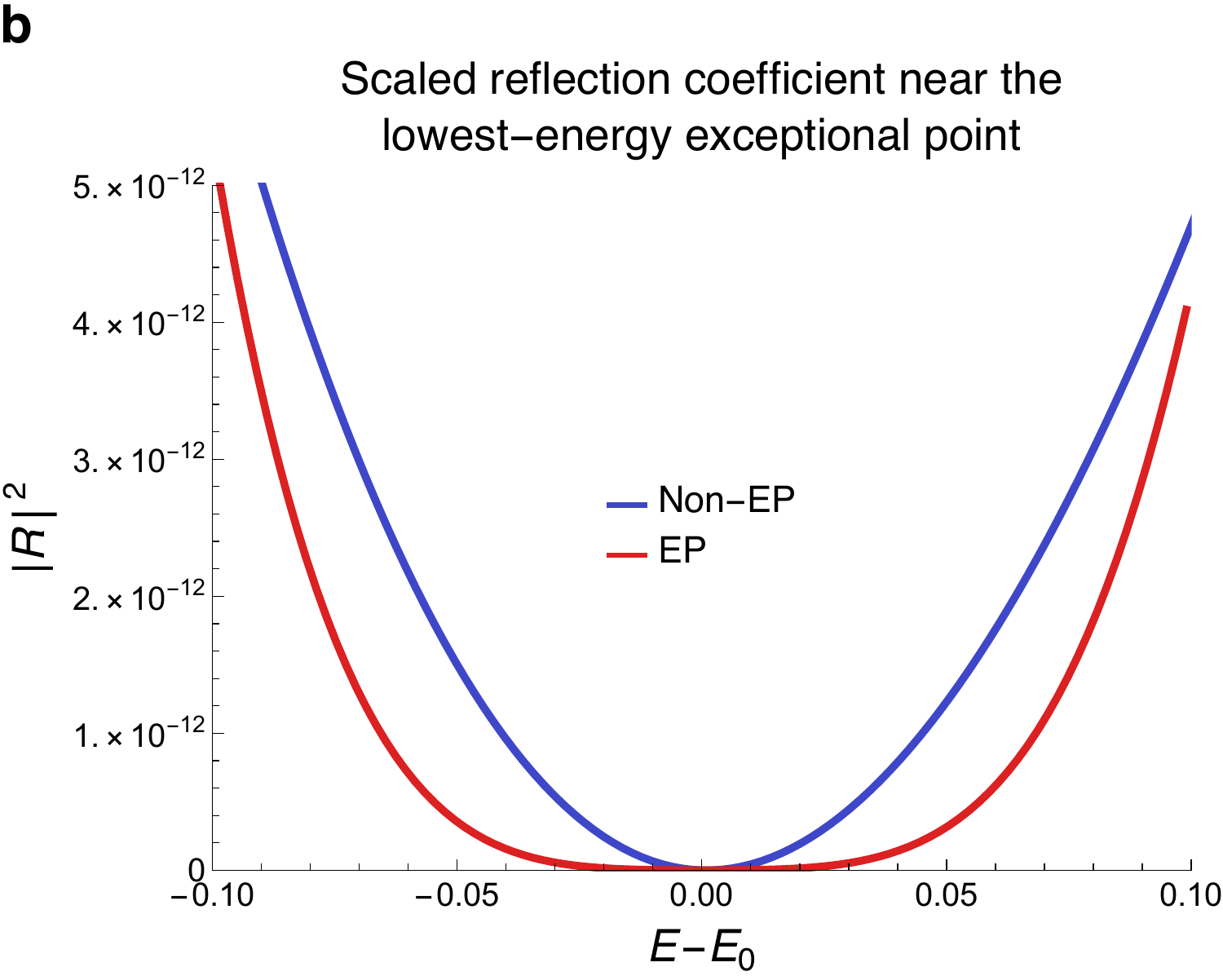}
\caption{(a)  Close-up of $\mathcal{PT}$-symmetry transitions at exceptional points in the RSM spectrum for $p \approx 3.44,E_\text{EP}=8.4$ and $p \approx 3.94,E_\text{EP}=20.2$. Left inset shows RSMs at $p=3.4$ below the lowest-energy exceptional point, which indicates a single real RSM accompanied by complex-conjugate RSMs in the broken $\mathcal{PT}$ phase prior to reaching the two lowest-energy exceptional points. Right inset demonstrates the unbroken to broken $\mathcal{PT}$  transition near the lowest-energy exceptional point, with a transition from complex-conjugate pairs of RSMs to a single real RSM. (b) The scattering near these EPs exhibits the anomalous quartic lineshape of the reflectance dip at $E = E_\text{EP}$ for a reflectionless EP, (red line, shown for $E_\text{EP} = 8.4$). In contrast an isolated RSM will have a quadratic behavior at the dip, as shown for $p=4.0,E_\text{RSM}=6.0$ (blue circle). Other signatures in the reflectance ratio transitions at at EPs are detailed in Appendix~\ref{sec:QuantumScatteringTruncation}. \label{fig:EP}}
\end{figure*}

The final signature of this quantum phase transition occurs for scattering at energies near the EPs (Fig.~\ref{fig:EP}(a)), at which two RSMs meet and then separate as they move into the complex plane. The general theory  \cite{Sweeney.2019.093901,Sweeney.2020.063511} predicts that the 
reflection coefficients at EPs will vanish like $(E-E_\text{EP})^4$, and not quadratically as for isolated RSMs. We confirm this behavior in Fig.~\ref{fig:EP}(b) and in additional related results in Appendix~\ref{sec:QuantumScatteringTruncation}. This straightforward measurement of the shape of the reflection dip away from and at an EP will provide a ``smoking-gun" confirmation of the presence of a quantum $\mathcal{PT}$ phase transition in this system.  The $\mathcal{PT}$ transition we identify here is driven by the increasing strength of above-barrier scattering as the variation of the potential becomes more rapid, and not by increasing loss/gain or differential loss, as in many prior studies.

To analyze further how the physics changes with the shape of the potential we employ an approach to identify the origin of above-barrier reflection, which we term the WKB {\em force} analysis \cite{Maitra.1996.4763,Soley.2021.L041301}. This method has been used previously to improve the accuracy of calculations of above-barrier reflection \cite{Maitra.1996.4763} and to reduce unphysical errors in quantum-mechanical simulations \cite{Soley.2021.L041301}; here we use the technique to determine the spatial origin of above-barrier reflection in these quantum potentials and the effect of truncation.

The method begins by observing that WKB wavefunctions, although approximate solutions of the Schr{\"o}dinger equation, cannot capture above-barrier reflection in 1D.  Starting with a positive momentum solution at $- \infty$, the local WKB momentum $p(x,E) \equiv \sqrt{2m(E-V(x))}$ can never change sign if $E > V_\text{max}$. In addition, it can be shown that given a potential $V(x)$, the WKB wavefunctions {\it exactly} satisfy a Schr{\"o}dinger equation to which, in addition to $V(x)$, one adds an energy-dependent potential correction $V_\text{WKB}(x,E)$ of order $\hbar^2$ of the form
\begin{align}
    V_\text{WKB}(x,&E)=\nonumber\\
    &-\hbar^2\left[\frac{5}{32m}\left(\frac{V^\prime(x)}{E-V(x)}\right)^2+\frac{V^{\prime\prime}(x)}{8m\left(E-V(x)\right)}\right].\nonumber
\end{align}
If we employ the reflectionless WKB states in the actual potential $V(x)$ to calculate above-barrier reflection using the distorted-wave Born approximation, the quantum reflection from the potential $V(x)$ at energy $E$ arises solely from the additional scattering induced by $V_\text{WKB}(x,E)$. Thus, $V_\text{WKB}(x,E)$ reveals the spatial location of above-barrier scattering; $V_\text{WKB}(x,E)$ is localized in a region around the potential maximum and vanishes far from the origin as $1/x^2$, even if $V(x)$ extends to infinity; thus, it can be used to analyze the infinite-length upside-down potentials $V(x) = - \lvert x \rvert^p$, as well as the truncated scattering potentials introduced here.

\begin{figure*}
\centering
\includegraphics[height=0.22\textheight]{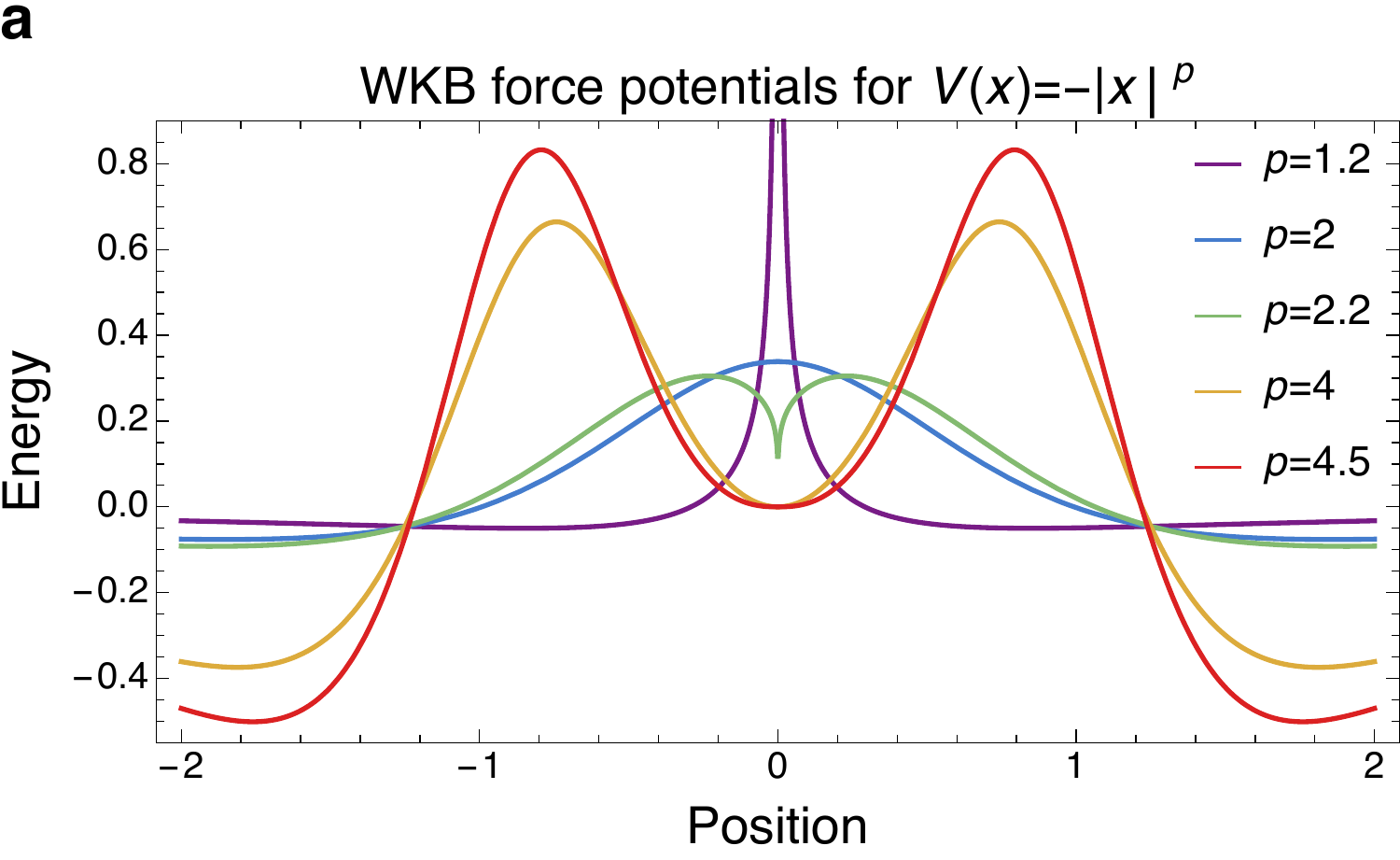}\,\includegraphics[height=0.22\textheight]{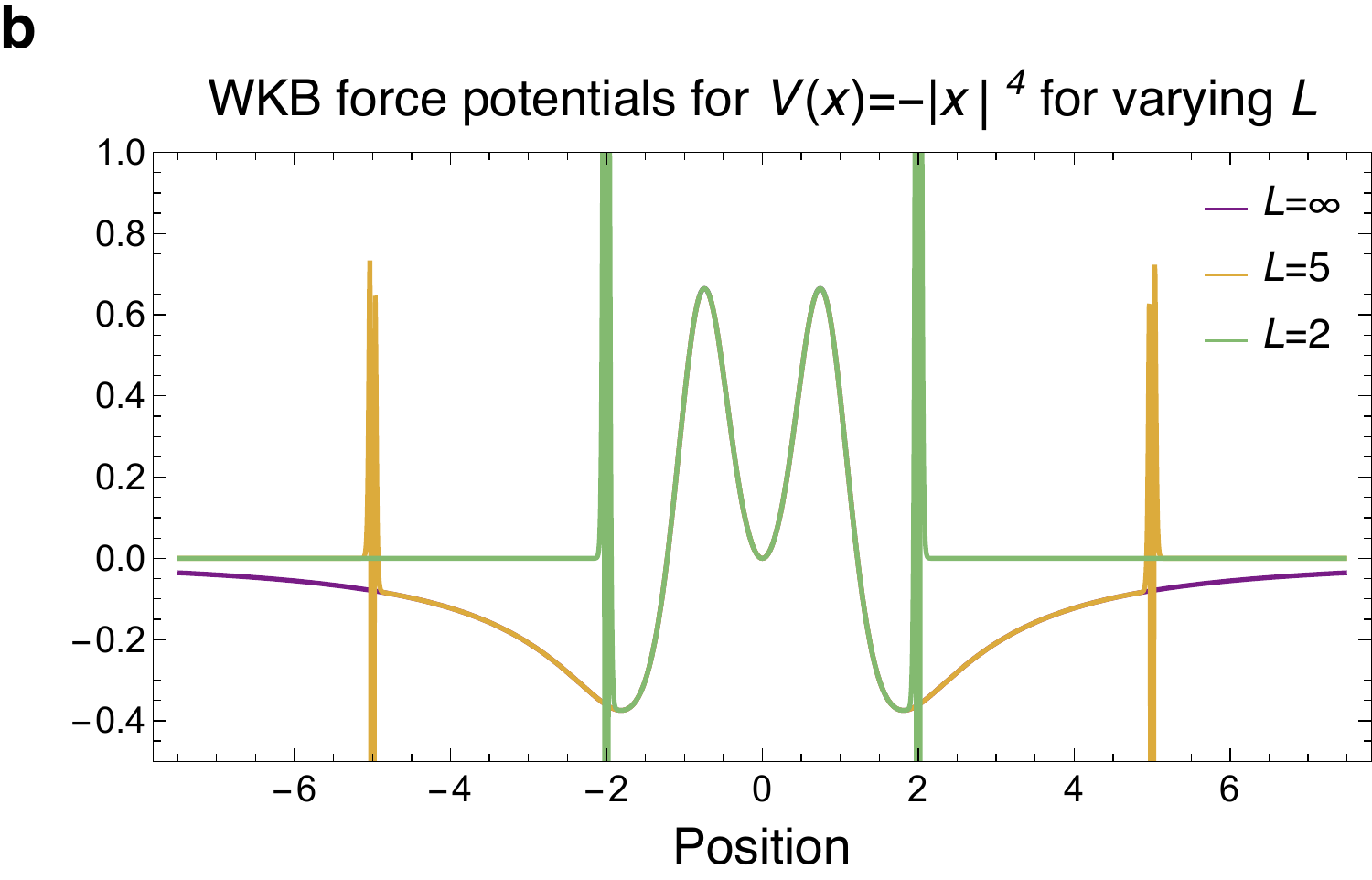}
\caption{WKB force analysis potentials at $E=1.477$ for (a) the $V(x)=-\lvert x\rvert ^p$ potential as a function of $p$ and (b) the truncated $V(x)=-\lvert x\rvert ^4$ and infinite-length $V(x)=-x^4$ potentials as a function of the truncation length $L$. The WKB force potential for varying $p$ values indicates a quantum phase transition in reflection at $p=2$ (solid blue line), in accordance with the quantum-scattering results and reflectionless scattering mode spectra.
The WKB force analysis potentials for the truncated potential with varying truncation lengths (solid colored lines) compared to the WKB force analysis potential for the infinite-length potential (dashed red line) suggest reflection due to truncation reduces as $L$ increases. Extended images are provided in Appendix~\ref{sec:WKBTruncation}.\label{fig:WKB}}
\end{figure*}

First, we calculate $V_\text{WKB}(x,E)$ for upside-down infinite-length potentials, to look for a signature of the quantum phase transition. We study the shape of $V_\text{WKB}(x,E,p)$ as $p$ is varied with $E=1.477$ (the ground state for $p=4$).
As shown in Fig.~\ref{fig:WKB}(a) and Appendix~\ref{sec:WKBTruncation}, there is a qualitative change in $V_\text{WKB}$ with $p$, and this change is insensitive to the value of $E$. For $p \leq 2$, $V_\text{WKB}$ has a single peak at the origin, which is divergent due to nonanalytic behavior at the origin for $p<2$. (Note that $V_\text{WKB}$ depends on $V^\prime$ and $V^{\prime \prime}$, at least one of which diverges for $p<2$.)  At $p=2, V(x)$ is analytic and for small $x$ the peak of $V_\text{WKB}$ is an inverted parabola.   However, for $p>2$ the peak splits into two maxima, symmetrically displaced from the origin, creating a small central well in $V_\text{WKB}$, and this well becomes deeper as $p \to 4$, where it creates an approximately parabolic ``trap." For $p>4$, the trap gets flatter at the center, making the scattering more localized at $x \approx \pm 1$, while $V(x)$ itself tends to a (semi-infinite in height) square potential well, which has an infinite set of reflectionless above-barrier resonances.

We now analyze $V_\text{WKB}$ for the {\em truncated} potential $V(x)=-\lvert x\rvert ^4$ at $E=1.477$.  In general, due to its local nature, $V_\text{WKB}$ is completely unchanged within the truncation length $L$, but the truncation does lead to a rapid spike at $L$, after which $V_\text{WKB}=0$ (as opposed to the $1/x^2$ decay for the infinite barrier).  The $V_\text{WKB}$ for the truncated $p=4$ potential with lengths $L=2$ and $5$ are shown in Fig.~\ref{fig:WKB}(b) and on a larger scale in Appendix~\ref{sec:WKBTruncation}. The size of the spikes in $V_\text{WKB}$ decreases as the truncation length is increased, indicating that the truncated potential reproduces the behavior of the infinite-length potentials at low energies.

In summary, the occurrence of a $\mathcal{PT}$ symmetry-breaking phase transition in the spectrum of above-barrier reflectionless quantum resonances of a class of real  quantum-scattering potentials brings this physics within the reach of standard cold trapped-atom experiments in engineered potentials, generated by techniques such as rapidly scanned lasers \cite{henderson2009experimental}, Digital Micromirror Devices (DMD) \cite{gauthier2016direct,navon2021quantum}, intensity masks \cite{scherer2007vortex}, and holographic methods  \cite{bergamini2004holographic,boyer2006dynamic,pasienski2008high,gaunt2012robust}. Since this spectrum is robust to smooth changes in the potential, these phenomena 
will be relatively insensitive to experimental nonidealities, such as  noise in the engineering of the power-law potential or introduction of additional polynomial terms (as we confirm in  Appendix~\ref{sec:Noise}). 

In addition, given the ubiquity of quantum reflection in near-threshold quantum systems, the occurrence of such phenomena may not be limited to the $V(x)=-\lvert x\rvert ^p$ potentials, but instead may be present in a wide range of quantum systems and possible quantum technologies.  Our results here should motivate a renewed search for $\mathcal{PT}$-symmetry behaviors in fundamental quantum systems and development of possible  $\mathcal{PT}$-symmetric quantum innovations in fields such as atomic-analogue lasing, quantum transport theory, and quantum sensing.

\section*{Methods}

\subsection*{Potential smoothing}

We determine the reflectionless scattering modes of the  $V(x)=-\lvert x \rvert^p$ potential in terms of the smoothed potentials
\begin{align}
V\left(x\right)&=-x^{4}f\left(x,w,L\right)-L^{4}f\left(-x,w,L\right),\nonumber\\
f\left(x,w,L\right) & =\frac{1}{1+e^{-w\left(x+L\right)}}+\frac{1}{1+e^{w\left(x-L\right)}}-1,\nonumber
\end{align}
where $f\left(x,w,L\right)$ is the smoothing function with sharpness parameter $w$ and truncation length $L$.  Note this class of potentials presents a distinct quantum-mechanical analysis from quantum-field-theoretical analysis of positive $V(x)=\lvert x\rvert ^p$ potentials \cite{bender1989new}, as these are upside-down and unbounded (before truncation) and feature no traditional bound states.

\subsection*{Quantum-scattering calculations}

Numerical quantum-scattering calculations are performed for the right-moving scattering solution to determine the reflectance ratio $\lvert R/T\rvert $ and subsequently the reflection coefficient $\lvert R\rvert ^2$ as detailed in Appendix~\ref{sec:QuantumScatteringTruncation}. The truncation length is accounted for implicitly both through the specification of the region in which the scattering wavefunction is determined and the Neumann boundary condition. For increased numerical stability, quantum-scattering calculations near exceptional points are determined via  explicit integration with Dormand-Prince coefficients.

\subsection*{RSM calculations}

According to RSM theory, under general conditions, for a finite-range potential, there exist eigensolutions in which the wave is incident in a given set of incoming scattering channels and exits through the complementary channels with zero reflection into the incoming channels.   These R-zero modes are found by considering the linear $N \times N$ scattering matrix, ${\bf S}$, which satisfies
\begin{equation}
    {\bf \beta}={\bf S}\left(E\right){\bf \alpha},\nonumber
\end{equation}
where ${\bf \alpha}$ is the vector of incoming wave amplitudes and ${\bf \beta}$ is the vector of outgoing wave amplitudes. Choosing $n <N $ incoming channels, let ${\bf R}_\text{in}\in\mathbb{C}^{n\times n}$ be the submatrix of the ${\bf S}$-matrix that connects the $n$ nonzero entries of the input wave vector ${\bf \alpha}$ to the $n$ corresponding entries of ${\bf \beta}$. An R-zero exists at (possibly complex) energy $E$ if there exists an $n$-component input ${\bf \alpha_\text{in}}$ at that energy for which all the corresponding $n$ outputs are zero,
\begin{equation}
{\bf R}_\text{in}\left(E\right){\bf \alpha}_\text{in}={\bf 0},\nonumber
\end{equation}
implying that $\det\left[{\bf R}_\text{in}(E)\right]=0$. The RSM theory demonstrates that under general conditions a countably infinite spectrum of R-zeros exists at complex energies for each choice of input space; these spectra are distinct, but similar in many ways, to the complex spectrum of resonances familiar in quantum scattering. As noted, if the system has $\mathcal{PT}$ symmetry, then these energies are either real or come in complex-conjugate pairs.  

RSMs are determined by imposing quadratic complex absorbing potentials to enforce boundary conditions for the right-moving solution and solving for the eigenenergies of the resulting non-Hermitian Schr{\"o}dinger equation via exact diagonalization. Here the truncation length $L$ and truncation rapidity $w$ are included explicitly. Both, as well as the strength and position of the complex absorbing potential and the choice of position space grid, are converged to yield the RSMs to the desired accuracy.

\subsection*{WKB force potential}

The WKB force potential is determined akin to the $W(x)$ potential of ref.~\cite{Maitra.1996.4763} or the WKB correction potential of ref.~\cite{Soley.2021.L041301}.  In contrast to previous works, the potential is considered alone without summation with either the original potential or additional terms ({\em e. g.}, the Coulomb potential) in order to directly pinpoint sources of quantum reflection in position space.

All calculations are evaluated in Wolfram Mathematica 12.3.1.0.

\section*{Acknowledgments} The authors thank E.~J.~Heller for stimulating conversations. MBS acknowledges financial support from the Yale Quantum Institute Postdoctoral Fellowship. ADS and CMB acknowledge financial support from the Simons Collaboration on Extreme Wave Phenomena Based on Symmetries.  CMB also acknowledges financial support from the Alexander von Humboldt Foundation and the UK Engineering and Physical Sciences Research Council (EPSRC) grant at King's College London.

\bibliography{PTSymmetry}

\begin{thebibliography}{50}%
\makeatletter
\providecommand \@ifxundefined [1]{%
 \@ifx{#1\undefined}
}%
\providecommand \@ifnum [1]{%
 \ifnum #1\expandafter \@firstoftwo
 \else \expandafter \@secondoftwo
 \fi
}%
\providecommand \@ifx [1]{%
 \ifx #1\expandafter \@firstoftwo
 \else \expandafter \@secondoftwo
 \fi
}%
\providecommand \natexlab [1]{#1}%
\providecommand \enquote  [1]{``#1''}%
\providecommand \bibnamefont  [1]{#1}%
\providecommand \bibfnamefont [1]{#1}%
\providecommand \citenamefont [1]{#1}%
\providecommand \href@noop [0]{\@secondoftwo}%
\providecommand \href [0]{\begingroup \@sanitize@url \@href}%
\providecommand \@href[1]{\@@startlink{#1}\@@href}%
\providecommand \@@href[1]{\endgroup#1\@@endlink}%
\providecommand \@sanitize@url [0]{\catcode `\\12\catcode `\$12\catcode
  `\&12\catcode `\#12\catcode `\^12\catcode `\_12\catcode `\%12\relax}%
\providecommand \@@startlink[1]{}%
\providecommand \@@endlink[0]{}%
\providecommand \url  [0]{\begingroup\@sanitize@url \@url }%
\providecommand \@url [1]{\endgroup\@href {#1}{\urlprefix }}%
\providecommand \urlprefix  [0]{URL }%
\providecommand \Eprint [0]{\href }%
\providecommand \doibase [0]{http://dx.doi.org/}%
\providecommand \selectlanguage [0]{\@gobble}%
\providecommand \bibinfo  [0]{\@secondoftwo}%
\providecommand \bibfield  [0]{\@secondoftwo}%
\providecommand \translation [1]{[#1]}%
\providecommand \BibitemOpen [0]{}%
\providecommand \bibitemStop [0]{}%
\providecommand \bibitemNoStop [0]{.\EOS\space}%
\providecommand \EOS [0]{\spacefactor3000\relax}%
\providecommand \BibitemShut  [1]{\csname bibitem#1\endcsname}%
\let\auto@bib@innerbib\@empty
\bibitem [{\citenamefont {Zhao}\ \emph {et~al.}(2010)\citenamefont {Zhao},
  \citenamefont {Schaden},\ and\ \citenamefont {Wu}}]{Zhao.2010.042903}%
  \BibitemOpen
  \bibfield  {author} {\bibinfo {author} {\bibfnamefont {K.~F.}\ \bibnamefont
  {Zhao}}, \bibinfo {author} {\bibfnamefont {M.}~\bibnamefont {Schaden}}, \
  and\ \bibinfo {author} {\bibfnamefont {Z.}~\bibnamefont {Wu}},\ }\href@noop
  {} {\bibfield  {journal} {\bibinfo  {journal} {Phys. Rev. A}\ }\textbf
  {\bibinfo {volume} {81}},\ \bibinfo {pages} {042903} (\bibinfo {year}
  {2010})}\BibitemShut {NoStop}%
\bibitem [{\citenamefont {Bittner}\ \emph {et~al.}(2012)\citenamefont
  {Bittner}, \citenamefont {Dietz}, \citenamefont {G{\"u}nther}, \citenamefont
  {Harney}, \citenamefont {Miski-Oglu}, \citenamefont {Richter},\ and\
  \citenamefont {Sch{\"a}fer}}]{Bittner.2012.024101}%
  \BibitemOpen
  \bibfield  {author} {\bibinfo {author} {\bibfnamefont {S.}~\bibnamefont
  {Bittner}}, \bibinfo {author} {\bibfnamefont {B.}~\bibnamefont {Dietz}},
  \bibinfo {author} {\bibfnamefont {U.}~\bibnamefont {G{\"u}nther}}, \bibinfo
  {author} {\bibfnamefont {H.~L.}\ \bibnamefont {Harney}}, \bibinfo {author}
  {\bibfnamefont {M.}~\bibnamefont {Miski-Oglu}}, \bibinfo {author}
  {\bibfnamefont {A.}~\bibnamefont {Richter}}, \ and\ \bibinfo {author}
  {\bibfnamefont {F.}~\bibnamefont {Sch{\"a}fer}},\ }\href@noop {} {\bibfield
  {journal} {\bibinfo  {journal} {Phys. Rev. Lett.}\ }\textbf {\bibinfo
  {volume} {108}},\ \bibinfo {pages} {024101} (\bibinfo {year}
  {2012})}\BibitemShut {NoStop}%
\bibitem [{\citenamefont {Zheng}\ \emph {et~al.}(2013)\citenamefont {Zheng},
  \citenamefont {Hao},\ and\ \citenamefont {Long}}]{Zheng.2013.20120053}%
  \BibitemOpen
  \bibfield  {author} {\bibinfo {author} {\bibfnamefont {C.}~\bibnamefont
  {Zheng}}, \bibinfo {author} {\bibfnamefont {L.}~\bibnamefont {Hao}}, \ and\
  \bibinfo {author} {\bibfnamefont {G.~L.}\ \bibnamefont {Long}},\ }\href@noop
  {} {\bibfield  {journal} {\bibinfo  {journal} {Philos. Trans. R. Soc. A}\
  }\textbf {\bibinfo {volume} {371}},\ \bibinfo {pages} {20120053} (\bibinfo
  {year} {2013})}\BibitemShut {NoStop}%
\bibitem [{\citenamefont {Guo}\ \emph {et~al.}(2009)\citenamefont {Guo},
  \citenamefont {Salamo}, \citenamefont {Duchesne}, \citenamefont {Morandotti},
  \citenamefont {Volatier-Ravat}, \citenamefont {Aimez}, \citenamefont
  {Siviloglou},\ and\ \citenamefont {Christodoulides}}]{Guo.2009.093902}%
  \BibitemOpen
  \bibfield  {author} {\bibinfo {author} {\bibfnamefont {A.}~\bibnamefont
  {Guo}}, \bibinfo {author} {\bibfnamefont {G.~J.}\ \bibnamefont {Salamo}},
  \bibinfo {author} {\bibfnamefont {D.}~\bibnamefont {Duchesne}}, \bibinfo
  {author} {\bibfnamefont {R.}~\bibnamefont {Morandotti}}, \bibinfo {author}
  {\bibfnamefont {M.}~\bibnamefont {Volatier-Ravat}}, \bibinfo {author}
  {\bibfnamefont {V.}~\bibnamefont {Aimez}}, \bibinfo {author} {\bibfnamefont
  {G.~A.}\ \bibnamefont {Siviloglou}}, \ and\ \bibinfo {author} {\bibfnamefont
  {D.~N.}\ \bibnamefont {Christodoulides}},\ }\href@noop {} {\bibfield
  {journal} {\bibinfo  {journal} {Phys. Rev. Lett.}\ }\textbf {\bibinfo
  {volume} {103}},\ \bibinfo {pages} {093902} (\bibinfo {year}
  {2009})}\BibitemShut {NoStop}%
\bibitem [{\citenamefont {R{\"u}ter}\ \emph {et~al.}(2010)\citenamefont
  {R{\"u}ter}, \citenamefont {Makris}, \citenamefont {El-Ganainy},
  \citenamefont {Christodoulides}, \citenamefont {Segev},\ and\ \citenamefont
  {Kip}}]{Ruter.2010.192}%
  \BibitemOpen
  \bibfield  {author} {\bibinfo {author} {\bibfnamefont {C.~E.}\ \bibnamefont
  {R{\"u}ter}}, \bibinfo {author} {\bibfnamefont {K.~G.}\ \bibnamefont
  {Makris}}, \bibinfo {author} {\bibfnamefont {R.}~\bibnamefont {El-Ganainy}},
  \bibinfo {author} {\bibfnamefont {D.~N.}\ \bibnamefont {Christodoulides}},
  \bibinfo {author} {\bibfnamefont {M.}~\bibnamefont {Segev}}, \ and\ \bibinfo
  {author} {\bibfnamefont {D.}~\bibnamefont {Kip}},\ }\href@noop {} {\bibfield
  {journal} {\bibinfo  {journal} {Nat. Phys.}\ }\textbf {\bibinfo {volume}
  {6}},\ \bibinfo {pages} {192} (\bibinfo {year} {2010})}\BibitemShut {NoStop}%
\bibitem [{\citenamefont {Feng}\ \emph {et~al.}(2011)\citenamefont {Feng},
  \citenamefont {Ayache}, \citenamefont {Huang}, \citenamefont {Xu},
  \citenamefont {Lu}, \citenamefont {Chen}, \citenamefont {Fainman},\ and\
  \citenamefont {Scherer}}]{Feng.2011.729}%
  \BibitemOpen
  \bibfield  {author} {\bibinfo {author} {\bibfnamefont {L.}~\bibnamefont
  {Feng}}, \bibinfo {author} {\bibfnamefont {M.}~\bibnamefont {Ayache}},
  \bibinfo {author} {\bibfnamefont {J.}~\bibnamefont {Huang}}, \bibinfo
  {author} {\bibfnamefont {Y.-L.}\ \bibnamefont {Xu}}, \bibinfo {author}
  {\bibfnamefont {M.-H.}\ \bibnamefont {Lu}}, \bibinfo {author} {\bibfnamefont
  {Y.-F.}\ \bibnamefont {Chen}}, \bibinfo {author} {\bibfnamefont
  {Y.}~\bibnamefont {Fainman}}, \ and\ \bibinfo {author} {\bibfnamefont
  {A.}~\bibnamefont {Scherer}},\ }\href@noop {} {\bibfield  {journal} {\bibinfo
   {journal} {Science}\ }\textbf {\bibinfo {volume} {333}},\ \bibinfo {pages}
  {729} (\bibinfo {year} {2011})}\BibitemShut {NoStop}%
\bibitem [{\citenamefont {Regensburger}\ \emph {et~al.}(2012)\citenamefont
  {Regensburger}, \citenamefont {Bersch}, \citenamefont {Miri}, \citenamefont
  {Onishchukov}, \citenamefont {Christodoulides},\ and\ \citenamefont
  {Peschel}}]{Regensburger.2012.167}%
  \BibitemOpen
  \bibfield  {author} {\bibinfo {author} {\bibfnamefont {A.}~\bibnamefont
  {Regensburger}}, \bibinfo {author} {\bibfnamefont {C.}~\bibnamefont
  {Bersch}}, \bibinfo {author} {\bibfnamefont {M.-A.}\ \bibnamefont {Miri}},
  \bibinfo {author} {\bibfnamefont {G.}~\bibnamefont {Onishchukov}}, \bibinfo
  {author} {\bibfnamefont {D.~N.}\ \bibnamefont {Christodoulides}}, \ and\
  \bibinfo {author} {\bibfnamefont {U.}~\bibnamefont {Peschel}},\ }\href@noop
  {} {\bibfield  {journal} {\bibinfo  {journal} {Nature}\ }\textbf {\bibinfo
  {volume} {488}},\ \bibinfo {pages} {167} (\bibinfo {year}
  {2012})}\BibitemShut {NoStop}%
\bibitem [{\citenamefont {Xiao}\ \emph {et~al.}(2021)\citenamefont {Xiao},
  \citenamefont {Deng}, \citenamefont {Wang}, \citenamefont {Wang},
  \citenamefont {Yi},\ and\ \citenamefont {Xue}}]{Xiao.2021.230402}%
  \BibitemOpen
  \bibfield  {author} {\bibinfo {author} {\bibfnamefont {L.}~\bibnamefont
  {Xiao}}, \bibinfo {author} {\bibfnamefont {T.}~\bibnamefont {Deng}}, \bibinfo
  {author} {\bibfnamefont {K.}~\bibnamefont {Wang}}, \bibinfo {author}
  {\bibfnamefont {Z.}~\bibnamefont {Wang}}, \bibinfo {author} {\bibfnamefont
  {W.}~\bibnamefont {Yi}}, \ and\ \bibinfo {author} {\bibfnamefont
  {P.}~\bibnamefont {Xue}},\ }\href@noop {} {\bibfield  {journal} {\bibinfo
  {journal} {Phys. Rev. Lett.}\ }\textbf {\bibinfo {volume} {126}},\ \bibinfo
  {pages} {230402} (\bibinfo {year} {2021})}\BibitemShut {NoStop}%
\bibitem [{\citenamefont {Peng}\ \emph {et~al.}(2014)\citenamefont {Peng},
  \citenamefont {{\"O}zdemir}, \citenamefont {Lei}, \citenamefont {Monifi},
  \citenamefont {Gianfreda}, \citenamefont {Long}, \citenamefont {Fan},
  \citenamefont {Nori}, \citenamefont {Bender},\ and\ \citenamefont
  {Yang}}]{Peng.2014.394}%
  \BibitemOpen
  \bibfield  {author} {\bibinfo {author} {\bibfnamefont {B.}~\bibnamefont
  {Peng}}, \bibinfo {author} {\bibfnamefont {{\c{S}}.~K.}\ \bibnamefont
  {{\"O}zdemir}}, \bibinfo {author} {\bibfnamefont {F.}~\bibnamefont {Lei}},
  \bibinfo {author} {\bibfnamefont {F.}~\bibnamefont {Monifi}}, \bibinfo
  {author} {\bibfnamefont {M.}~\bibnamefont {Gianfreda}}, \bibinfo {author}
  {\bibfnamefont {G.~L.}\ \bibnamefont {Long}}, \bibinfo {author}
  {\bibfnamefont {S.}~\bibnamefont {Fan}}, \bibinfo {author} {\bibfnamefont
  {F.}~\bibnamefont {Nori}}, \bibinfo {author} {\bibfnamefont {C.~M.}\
  \bibnamefont {Bender}}, \ and\ \bibinfo {author} {\bibfnamefont
  {L.}~\bibnamefont {Yang}},\ }\href@noop {} {\bibfield  {journal} {\bibinfo
  {journal} {Nat. Phys.}\ }\textbf {\bibinfo {volume} {10}},\ \bibinfo {pages}
  {394} (\bibinfo {year} {2014})}\BibitemShut {NoStop}%
\bibitem [{\citenamefont {Shi}\ \emph {et~al.}(2016)\citenamefont {Shi},
  \citenamefont {Dubois}, \citenamefont {Chen}, \citenamefont {Cheng},
  \citenamefont {Ramezani}, \citenamefont {Wang},\ and\ \citenamefont
  {Zhang}}]{Shi.2016.1}%
  \BibitemOpen
  \bibfield  {author} {\bibinfo {author} {\bibfnamefont {C.}~\bibnamefont
  {Shi}}, \bibinfo {author} {\bibfnamefont {M.}~\bibnamefont {Dubois}},
  \bibinfo {author} {\bibfnamefont {Y.}~\bibnamefont {Chen}}, \bibinfo {author}
  {\bibfnamefont {L.}~\bibnamefont {Cheng}}, \bibinfo {author} {\bibfnamefont
  {H.}~\bibnamefont {Ramezani}}, \bibinfo {author} {\bibfnamefont
  {Y.}~\bibnamefont {Wang}}, \ and\ \bibinfo {author} {\bibfnamefont
  {X.}~\bibnamefont {Zhang}},\ }\href@noop {} {\bibfield  {journal} {\bibinfo
  {journal} {Nat. Commun.}\ }\textbf {\bibinfo {volume} {7}},\ \bibinfo {pages}
  {1} (\bibinfo {year} {2016})}\BibitemShut {NoStop}%
\bibitem [{\citenamefont {Aur{\'e}gan}\ and\ \citenamefont
  {Pagneux}(2017)}]{Auregan.2017.174301}%
  \BibitemOpen
  \bibfield  {author} {\bibinfo {author} {\bibfnamefont {Y.}~\bibnamefont
  {Aur{\'e}gan}}\ and\ \bibinfo {author} {\bibfnamefont {V.}~\bibnamefont
  {Pagneux}},\ }\href@noop {} {\bibfield  {journal} {\bibinfo  {journal} {Phys.
  Rev. Lett.}\ }\textbf {\bibinfo {volume} {118}},\ \bibinfo {pages} {174301}
  (\bibinfo {year} {2017})}\BibitemShut {NoStop}%
\bibitem [{\citenamefont {Chtchelkatchev}\ \emph {et~al.}(2012)\citenamefont
  {Chtchelkatchev}, \citenamefont {Golubov}, \citenamefont {Baturina},\ and\
  \citenamefont {Vinokur}}]{Chtchelkatchev.2012.150405}%
  \BibitemOpen
  \bibfield  {author} {\bibinfo {author} {\bibfnamefont {N.~M.}\ \bibnamefont
  {Chtchelkatchev}}, \bibinfo {author} {\bibfnamefont {A.~A.}\ \bibnamefont
  {Golubov}}, \bibinfo {author} {\bibfnamefont {T.~I.}\ \bibnamefont
  {Baturina}}, \ and\ \bibinfo {author} {\bibfnamefont {V.~M.}\ \bibnamefont
  {Vinokur}},\ }\href@noop {} {\bibfield  {journal} {\bibinfo  {journal} {Phys.
  Rev. Lett.}\ }\textbf {\bibinfo {volume} {109}},\ \bibinfo {pages} {150405}
  (\bibinfo {year} {2012})}\BibitemShut {NoStop}%
\bibitem [{\citenamefont {Schindler}\ \emph {et~al.}(2011)\citenamefont
  {Schindler}, \citenamefont {Li}, \citenamefont {Zheng}, \citenamefont
  {Ellis},\ and\ \citenamefont {Kottos}}]{Schindler.2011.040101}%
  \BibitemOpen
  \bibfield  {author} {\bibinfo {author} {\bibfnamefont {J.}~\bibnamefont
  {Schindler}}, \bibinfo {author} {\bibfnamefont {A.}~\bibnamefont {Li}},
  \bibinfo {author} {\bibfnamefont {M.~C.}\ \bibnamefont {Zheng}}, \bibinfo
  {author} {\bibfnamefont {F.~M.}\ \bibnamefont {Ellis}}, \ and\ \bibinfo
  {author} {\bibfnamefont {T.}~\bibnamefont {Kottos}},\ }\href@noop {}
  {\bibfield  {journal} {\bibinfo  {journal} {Phys. Rev. A}\ }\textbf {\bibinfo
  {volume} {84}},\ \bibinfo {pages} {040101} (\bibinfo {year}
  {2011})}\BibitemShut {NoStop}%
\bibitem [{\citenamefont {Bender}\ \emph
  {et~al.}(2013{\natexlab{a}})\citenamefont {Bender}, \citenamefont {Factor},
  \citenamefont {Bodyfelt}, \citenamefont {Ramezani}, \citenamefont
  {Christodoulides}, \citenamefont {Ellis},\ and\ \citenamefont
  {Kottos}}]{Bender.2013.234101}%
  \BibitemOpen
  \bibfield  {author} {\bibinfo {author} {\bibfnamefont {N.}~\bibnamefont
  {Bender}}, \bibinfo {author} {\bibfnamefont {S.}~\bibnamefont {Factor}},
  \bibinfo {author} {\bibfnamefont {J.~D.}\ \bibnamefont {Bodyfelt}}, \bibinfo
  {author} {\bibfnamefont {H.}~\bibnamefont {Ramezani}}, \bibinfo {author}
  {\bibfnamefont {D.~N.}\ \bibnamefont {Christodoulides}}, \bibinfo {author}
  {\bibfnamefont {F.~M.}\ \bibnamefont {Ellis}}, \ and\ \bibinfo {author}
  {\bibfnamefont {T.}~\bibnamefont {Kottos}},\ }\href@noop {} {\bibfield
  {journal} {\bibinfo  {journal} {Phys. Rev. Lett.}\ }\textbf {\bibinfo
  {volume} {110}},\ \bibinfo {pages} {234101} (\bibinfo {year}
  {2013}{\natexlab{a}})}\BibitemShut {NoStop}%
\bibitem [{\citenamefont {Cao}\ \emph {et~al.}(2022)\citenamefont {Cao},
  \citenamefont {Wang}, \citenamefont {Chen}, \citenamefont {Hu}, \citenamefont
  {Wang}, \citenamefont {Yang},\ and\ \citenamefont {Zhang}}]{Cao.2022.1}%
  \BibitemOpen
  \bibfield  {author} {\bibinfo {author} {\bibfnamefont {W.}~\bibnamefont
  {Cao}}, \bibinfo {author} {\bibfnamefont {C.}~\bibnamefont {Wang}}, \bibinfo
  {author} {\bibfnamefont {W.}~\bibnamefont {Chen}}, \bibinfo {author}
  {\bibfnamefont {S.}~\bibnamefont {Hu}}, \bibinfo {author} {\bibfnamefont
  {H.}~\bibnamefont {Wang}}, \bibinfo {author} {\bibfnamefont {L.}~\bibnamefont
  {Yang}}, \ and\ \bibinfo {author} {\bibfnamefont {X.}~\bibnamefont {Zhang}},\
  }\href@noop {} {\bibfield  {journal} {\bibinfo  {journal} {Nature
  Nanotechnol.}\ ,\ \bibinfo {pages} {1}} (\bibinfo {year} {2022})}\BibitemShut
  {NoStop}%
\bibitem [{\citenamefont {Bender}\ \emph
  {et~al.}(2013{\natexlab{b}})\citenamefont {Bender}, \citenamefont {Berntson},
  \citenamefont {Parker},\ and\ \citenamefont {Samuel}}]{Bender.2013.173}%
  \BibitemOpen
  \bibfield  {author} {\bibinfo {author} {\bibfnamefont {C.~M.}\ \bibnamefont
  {Bender}}, \bibinfo {author} {\bibfnamefont {B.~K.}\ \bibnamefont
  {Berntson}}, \bibinfo {author} {\bibfnamefont {D.}~\bibnamefont {Parker}}, \
  and\ \bibinfo {author} {\bibfnamefont {E.}~\bibnamefont {Samuel}},\
  }\href@noop {} {\bibfield  {journal} {\bibinfo  {journal} {Am. J. Phys.}\
  }\textbf {\bibinfo {volume} {81}},\ \bibinfo {pages} {173} (\bibinfo {year}
  {2013}{\natexlab{b}})}\BibitemShut {NoStop}%
\bibitem [{\citenamefont {Feng}\ \emph {et~al.}(2014)\citenamefont {Feng},
  \citenamefont {Wong}, \citenamefont {Ma}, \citenamefont {Wang},\ and\
  \citenamefont {Zhang}}]{Feng.2014.972}%
  \BibitemOpen
  \bibfield  {author} {\bibinfo {author} {\bibfnamefont {L.}~\bibnamefont
  {Feng}}, \bibinfo {author} {\bibfnamefont {Z.~J.}\ \bibnamefont {Wong}},
  \bibinfo {author} {\bibfnamefont {R.-M.}\ \bibnamefont {Ma}}, \bibinfo
  {author} {\bibfnamefont {Y.}~\bibnamefont {Wang}}, \ and\ \bibinfo {author}
  {\bibfnamefont {X.}~\bibnamefont {Zhang}},\ }\href@noop {} {\bibfield
  {journal} {\bibinfo  {journal} {Science}\ }\textbf {\bibinfo {volume}
  {346}},\ \bibinfo {pages} {972} (\bibinfo {year} {2014})}\BibitemShut
  {NoStop}%
\bibitem [{\citenamefont {Hodaei}\ \emph {et~al.}(2014)\citenamefont {Hodaei},
  \citenamefont {Miri}, \citenamefont {Heinrich}, \citenamefont
  {Christodoulides},\ and\ \citenamefont {Khajavikhan}}]{Hodaei.2014.975}%
  \BibitemOpen
  \bibfield  {author} {\bibinfo {author} {\bibfnamefont {H.}~\bibnamefont
  {Hodaei}}, \bibinfo {author} {\bibfnamefont {M.-A.}\ \bibnamefont {Miri}},
  \bibinfo {author} {\bibfnamefont {M.}~\bibnamefont {Heinrich}}, \bibinfo
  {author} {\bibfnamefont {D.~N.}\ \bibnamefont {Christodoulides}}, \ and\
  \bibinfo {author} {\bibfnamefont {M.}~\bibnamefont {Khajavikhan}},\
  }\href@noop {} {\bibfield  {journal} {\bibinfo  {journal} {Science}\ }\textbf
  {\bibinfo {volume} {346}},\ \bibinfo {pages} {975} (\bibinfo {year}
  {2014})}\BibitemShut {NoStop}%
\bibitem [{\citenamefont {Chen}\ and\ \citenamefont {Jung}(2016)}]{chen2016p}%
  \BibitemOpen
  \bibfield  {author} {\bibinfo {author} {\bibfnamefont {P.-Y.}\ \bibnamefont
  {Chen}}\ and\ \bibinfo {author} {\bibfnamefont {J.}~\bibnamefont {Jung}},\
  }\href@noop {} {\bibfield  {journal} {\bibinfo  {journal} {Phys. Rev. Appl.}\
  }\textbf {\bibinfo {volume} {5}},\ \bibinfo {pages} {064018} (\bibinfo {year}
  {2016})}\BibitemShut {NoStop}%
\bibitem [{\citenamefont {Liu}\ \emph {et~al.}(2016)\citenamefont {Liu},
  \citenamefont {Zhang}, \citenamefont {{\"O}zdemir}, \citenamefont {Peng},
  \citenamefont {Jing}, \citenamefont {L{\"u}}, \citenamefont {Li},
  \citenamefont {Yang}, \citenamefont {Nori},\ and\ \citenamefont
  {Liu}}]{liu2016metrology}%
  \BibitemOpen
  \bibfield  {author} {\bibinfo {author} {\bibfnamefont {Z.-P.}\ \bibnamefont
  {Liu}}, \bibinfo {author} {\bibfnamefont {J.}~\bibnamefont {Zhang}}, \bibinfo
  {author} {\bibfnamefont {{\c{S}}.~K.}\ \bibnamefont {{\"O}zdemir}}, \bibinfo
  {author} {\bibfnamefont {B.}~\bibnamefont {Peng}}, \bibinfo {author}
  {\bibfnamefont {H.}~\bibnamefont {Jing}}, \bibinfo {author} {\bibfnamefont
  {X.-Y.}\ \bibnamefont {L{\"u}}}, \bibinfo {author} {\bibfnamefont {C.-W.}\
  \bibnamefont {Li}}, \bibinfo {author} {\bibfnamefont {L.}~\bibnamefont
  {Yang}}, \bibinfo {author} {\bibfnamefont {F.}~\bibnamefont {Nori}}, \ and\
  \bibinfo {author} {\bibfnamefont {Y.}~\bibnamefont {Liu}},\ }\href@noop {}
  {\bibfield  {journal} {\bibinfo  {journal} {Phys. Rev. Lett.}\ }\textbf
  {\bibinfo {volume} {117}},\ \bibinfo {pages} {110802} (\bibinfo {year}
  {2016})}\BibitemShut {NoStop}%
\bibitem [{\citenamefont {Assawaworrarit}\ \emph {et~al.}(2017)\citenamefont
  {Assawaworrarit}, \citenamefont {Yu},\ and\ \citenamefont
  {Fan}}]{Assawaworrarit.2017.387}%
  \BibitemOpen
  \bibfield  {author} {\bibinfo {author} {\bibfnamefont {S.}~\bibnamefont
  {Assawaworrarit}}, \bibinfo {author} {\bibfnamefont {X.}~\bibnamefont {Yu}},
  \ and\ \bibinfo {author} {\bibfnamefont {S.}~\bibnamefont {Fan}},\
  }\href@noop {} {\bibfield  {journal} {\bibinfo  {journal} {Nature}\ }\textbf
  {\bibinfo {volume} {546}},\ \bibinfo {pages} {387} (\bibinfo {year}
  {2017})}\BibitemShut {NoStop}%
\bibitem [{\citenamefont {Assawaworrarit}\ and\ \citenamefont
  {Fan}(2020)}]{Assawaworrarit.2020.273}%
  \BibitemOpen
  \bibfield  {author} {\bibinfo {author} {\bibfnamefont {S.}~\bibnamefont
  {Assawaworrarit}}\ and\ \bibinfo {author} {\bibfnamefont {S.}~\bibnamefont
  {Fan}},\ }\href@noop {} {\bibfield  {journal} {\bibinfo  {journal} {Nat.
  Electron.}\ }\textbf {\bibinfo {volume} {3}},\ \bibinfo {pages} {273}
  (\bibinfo {year} {2020})}\BibitemShut {NoStop}%
\bibitem [{\citenamefont {Bender}\ and\ \citenamefont
  {Boettcher}(1998)}]{Bender.1998.5243}%
  \BibitemOpen
  \bibfield  {author} {\bibinfo {author} {\bibfnamefont {C.~M.}\ \bibnamefont
  {Bender}}\ and\ \bibinfo {author} {\bibfnamefont {S.}~\bibnamefont
  {Boettcher}},\ }\href@noop {} {\bibfield  {journal} {\bibinfo  {journal}
  {Phys. Rev. Lett.}\ }\textbf {\bibinfo {volume} {80}},\ \bibinfo {pages}
  {5243} (\bibinfo {year} {1998})}\BibitemShut {NoStop}%
\bibitem [{\citenamefont {Bender}(2007)}]{Bender.2007.947}%
  \BibitemOpen
  \bibfield  {author} {\bibinfo {author} {\bibfnamefont {C.~M.}\ \bibnamefont
  {Bender}},\ }\href@noop {} {\bibfield  {journal} {\bibinfo  {journal} {Rep.
  Prog. Phys.}\ }\textbf {\bibinfo {volume} {70}},\ \bibinfo {pages} {947}
  (\bibinfo {year} {2007})}\BibitemShut {NoStop}%
\bibitem [{\citenamefont {Ahmed}\ \emph {et~al.}(2005)\citenamefont {Ahmed},
  \citenamefont {Bender},\ and\ \citenamefont
  {Berry}}]{ahmed2005reflectionless}%
  \BibitemOpen
  \bibfield  {author} {\bibinfo {author} {\bibfnamefont {Z.}~\bibnamefont
  {Ahmed}}, \bibinfo {author} {\bibfnamefont {C.~M.}\ \bibnamefont {Bender}}, \
  and\ \bibinfo {author} {\bibfnamefont {M.~V.}\ \bibnamefont {Berry}},\
  }\href@noop {} {\bibfield  {journal} {\bibinfo  {journal} {J. Phys. A: Math.
  Gen.}\ }\textbf {\bibinfo {volume} {38}},\ \bibinfo {pages} {L627} (\bibinfo
  {year} {2005})}\BibitemShut {NoStop}%
\bibitem [{\citenamefont {Bender}\ and\ \citenamefont
  {Gianfreda}(2018)}]{Bender.2018.052118}%
  \BibitemOpen
  \bibfield  {author} {\bibinfo {author} {\bibfnamefont {C.~M.}\ \bibnamefont
  {Bender}}\ and\ \bibinfo {author} {\bibfnamefont {M.}~\bibnamefont
  {Gianfreda}},\ }\href@noop {} {\bibfield  {journal} {\bibinfo  {journal}
  {Phys. Rev. A}\ }\textbf {\bibinfo {volume} {5}},\ \bibinfo {pages} {052118}
  (\bibinfo {year} {2018})}\BibitemShut {NoStop}%
\bibitem [{\citenamefont {Sweeney}\ \emph {et~al.}(2020)\citenamefont
  {Sweeney}, \citenamefont {Hsu},\ and\ \citenamefont
  {Stone}}]{Sweeney.2020.063511}%
  \BibitemOpen
  \bibfield  {author} {\bibinfo {author} {\bibfnamefont {W.~R.}\ \bibnamefont
  {Sweeney}}, \bibinfo {author} {\bibfnamefont {C.~W.}\ \bibnamefont {Hsu}}, \
  and\ \bibinfo {author} {\bibfnamefont {A.~D.}\ \bibnamefont {Stone}},\
  }\href@noop {} {\bibfield  {journal} {\bibinfo  {journal} {Phys. Rev. A}\
  }\textbf {\bibinfo {volume} {102}},\ \bibinfo {pages} {063511} (\bibinfo
  {year} {2020})}\BibitemShut {NoStop}%
\bibitem [{\citenamefont {Stone}\ \emph {et~al.}(2021)\citenamefont {Stone},
  \citenamefont {Sweeney}, \citenamefont {Hsu}, \citenamefont {Wisal},\ and\
  \citenamefont {Wang}}]{Stone.2020.343}%
  \BibitemOpen
  \bibfield  {author} {\bibinfo {author} {\bibfnamefont {A.~D.}\ \bibnamefont
  {Stone}}, \bibinfo {author} {\bibfnamefont {W.~R.}\ \bibnamefont {Sweeney}},
  \bibinfo {author} {\bibfnamefont {C.~W.}\ \bibnamefont {Hsu}}, \bibinfo
  {author} {\bibfnamefont {K.}~\bibnamefont {Wisal}}, \ and\ \bibinfo {author}
  {\bibfnamefont {Z.}~\bibnamefont {Wang}},\ }\href@noop {} {\bibfield
  {journal} {\bibinfo  {journal} {Nanophotonics}\ }\textbf {\bibinfo {volume}
  {10}},\ \bibinfo {pages} {343} (\bibinfo {year} {2021})}\BibitemShut
  {NoStop}%
\bibitem [{\citenamefont {Moiseyev}(1998)}]{moiseyev1998quantum}%
  \BibitemOpen
  \bibfield  {author} {\bibinfo {author} {\bibfnamefont {N.}~\bibnamefont
  {Moiseyev}},\ }\href@noop {} {\bibfield  {journal} {\bibinfo  {journal}
  {Phys. Rep.}\ }\textbf {\bibinfo {volume} {302}},\ \bibinfo {pages} {212}
  (\bibinfo {year} {1998})}\BibitemShut {NoStop}%
\bibitem [{\citenamefont {Muga}\ \emph {et~al.}(2004)\citenamefont {Muga},
  \citenamefont {Palao}, \citenamefont {Navarro},\ and\ \citenamefont
  {Egusquiza}}]{muga2004complex}%
  \BibitemOpen
  \bibfield  {author} {\bibinfo {author} {\bibfnamefont {J.}~\bibnamefont
  {Muga}}, \bibinfo {author} {\bibfnamefont {J.}~\bibnamefont {Palao}},
  \bibinfo {author} {\bibfnamefont {B.}~\bibnamefont {Navarro}}, \ and\
  \bibinfo {author} {\bibfnamefont {I.}~\bibnamefont {Egusquiza}},\ }\href@noop
  {} {\bibfield  {journal} {\bibinfo  {journal} {Phys. Rep.}\ }\textbf
  {\bibinfo {volume} {395}},\ \bibinfo {pages} {357} (\bibinfo {year}
  {2004})}\BibitemShut {NoStop}%
\bibitem [{\citenamefont {Kemble}(1935)}]{kemble1935contribution}%
  \BibitemOpen
  \bibfield  {author} {\bibinfo {author} {\bibfnamefont {E.~C.}\ \bibnamefont
  {Kemble}},\ }\href@noop {} {\bibfield  {journal} {\bibinfo  {journal} {Phys.
  Rev.}\ }\textbf {\bibinfo {volume} {48}},\ \bibinfo {pages} {549} (\bibinfo
  {year} {1935})}\BibitemShut {NoStop}%
\bibitem [{\citenamefont {Sweeney}\ \emph {et~al.}(2019)\citenamefont
  {Sweeney}, \citenamefont {Hsu}, \citenamefont {Rotter},\ and\ \citenamefont
  {Stone}}]{Sweeney.2019.093901}%
  \BibitemOpen
  \bibfield  {author} {\bibinfo {author} {\bibfnamefont {W.~R.}\ \bibnamefont
  {Sweeney}}, \bibinfo {author} {\bibfnamefont {C.~W.}\ \bibnamefont {Hsu}},
  \bibinfo {author} {\bibfnamefont {S.}~\bibnamefont {Rotter}}, \ and\ \bibinfo
  {author} {\bibfnamefont {A.~D.}\ \bibnamefont {Stone}},\ }\href@noop {}
  {\bibfield  {journal} {\bibinfo  {journal} {Phys. Rev. Lett.}\ }\textbf
  {\bibinfo {volume} {122}},\ \bibinfo {pages} {093901} (\bibinfo {year}
  {2019})}\BibitemShut {NoStop}%
\bibitem [{\citenamefont {Maitra}\ and\ \citenamefont
  {Heller}(1996)}]{Maitra.1996.4763}%
  \BibitemOpen
  \bibfield  {author} {\bibinfo {author} {\bibfnamefont {N.~T.}\ \bibnamefont
  {Maitra}}\ and\ \bibinfo {author} {\bibfnamefont {E.~J.}\ \bibnamefont
  {Heller}},\ }\href@noop {} {\bibfield  {journal} {\bibinfo  {journal} {Phys.
  Rev. A}\ }\textbf {\bibinfo {volume} {54}},\ \bibinfo {pages} {4763}
  (\bibinfo {year} {1996})}\BibitemShut {NoStop}%
\bibitem [{\citenamefont {Soley}\ \emph {et~al.}(2021)\citenamefont {Soley},
  \citenamefont {Avanaki},\ and\ \citenamefont {Heller}}]{Soley.2021.L041301}%
  \BibitemOpen
  \bibfield  {author} {\bibinfo {author} {\bibfnamefont {M.~B.}\ \bibnamefont
  {Soley}}, \bibinfo {author} {\bibfnamefont {K.~N.}\ \bibnamefont {Avanaki}},
  \ and\ \bibinfo {author} {\bibfnamefont {E.~J.}\ \bibnamefont {Heller}},\
  }\href@noop {} {\bibfield  {journal} {\bibinfo  {journal} {Phys. Rev. A}\
  }\textbf {\bibinfo {volume} {103}},\ \bibinfo {pages} {L041301} (\bibinfo
  {year} {2021})}\BibitemShut {NoStop}%
\bibitem [{\citenamefont {Henderson}\ \emph {et~al.}(2009)\citenamefont
  {Henderson}, \citenamefont {Ryu}, \citenamefont {MacCormick},\ and\
  \citenamefont {Boshier}}]{henderson2009experimental}%
  \BibitemOpen
  \bibfield  {author} {\bibinfo {author} {\bibfnamefont {K.}~\bibnamefont
  {Henderson}}, \bibinfo {author} {\bibfnamefont {C.}~\bibnamefont {Ryu}},
  \bibinfo {author} {\bibfnamefont {C.}~\bibnamefont {MacCormick}}, \ and\
  \bibinfo {author} {\bibfnamefont {M.}~\bibnamefont {Boshier}},\ }\href@noop
  {} {\bibfield  {journal} {\bibinfo  {journal} {New J. Phys.}\ }\textbf
  {\bibinfo {volume} {11}},\ \bibinfo {pages} {043030} (\bibinfo {year}
  {2009})}\BibitemShut {NoStop}%
\bibitem [{\citenamefont {Gauthier}\ \emph {et~al.}(2016)\citenamefont
  {Gauthier}, \citenamefont {Lenton}, \citenamefont {Parry}, \citenamefont
  {Baker}, \citenamefont {Davis}, \citenamefont {Rubinsztein-Dunlop},\ and\
  \citenamefont {Neely}}]{gauthier2016direct}%
  \BibitemOpen
  \bibfield  {author} {\bibinfo {author} {\bibfnamefont {G.}~\bibnamefont
  {Gauthier}}, \bibinfo {author} {\bibfnamefont {I.}~\bibnamefont {Lenton}},
  \bibinfo {author} {\bibfnamefont {N.~M.}\ \bibnamefont {Parry}}, \bibinfo
  {author} {\bibfnamefont {M.}~\bibnamefont {Baker}}, \bibinfo {author}
  {\bibfnamefont {M.}~\bibnamefont {Davis}}, \bibinfo {author} {\bibfnamefont
  {H.}~\bibnamefont {Rubinsztein-Dunlop}}, \ and\ \bibinfo {author}
  {\bibfnamefont {T.}~\bibnamefont {Neely}},\ }\href@noop {} {\bibfield
  {journal} {\bibinfo  {journal} {Optica}\ }\textbf {\bibinfo {volume} {3}},\
  \bibinfo {pages} {1136} (\bibinfo {year} {2016})}\BibitemShut {NoStop}%
\bibitem [{\citenamefont {Navon}\ \emph {et~al.}(2021)\citenamefont {Navon},
  \citenamefont {Smith},\ and\ \citenamefont {Hadzibabic}}]{navon2021quantum}%
  \BibitemOpen
  \bibfield  {author} {\bibinfo {author} {\bibfnamefont {N.}~\bibnamefont
  {Navon}}, \bibinfo {author} {\bibfnamefont {R.~P.}\ \bibnamefont {Smith}}, \
  and\ \bibinfo {author} {\bibfnamefont {Z.}~\bibnamefont {Hadzibabic}},\
  }\href@noop {} {\bibfield  {journal} {\bibinfo  {journal} {Nat. Phys.}\ ,\
  \bibinfo {pages} {1}} (\bibinfo {year} {2021})}\BibitemShut {NoStop}%
\bibitem [{\citenamefont {Scherer}\ \emph {et~al.}(2007)\citenamefont
  {Scherer}, \citenamefont {Weiler}, \citenamefont {Neely},\ and\ \citenamefont
  {Anderson}}]{scherer2007vortex}%
  \BibitemOpen
  \bibfield  {author} {\bibinfo {author} {\bibfnamefont {D.~R.}\ \bibnamefont
  {Scherer}}, \bibinfo {author} {\bibfnamefont {C.~N.}\ \bibnamefont {Weiler}},
  \bibinfo {author} {\bibfnamefont {T.~W.}\ \bibnamefont {Neely}}, \ and\
  \bibinfo {author} {\bibfnamefont {B.~P.}\ \bibnamefont {Anderson}},\
  }\href@noop {} {\bibfield  {journal} {\bibinfo  {journal} {Phys. Rev. Lett.}\
  }\textbf {\bibinfo {volume} {98}},\ \bibinfo {pages} {110402} (\bibinfo
  {year} {2007})}\BibitemShut {NoStop}%
\bibitem [{\citenamefont {Bergamini}\ \emph {et~al.}(2004)\citenamefont
  {Bergamini}, \citenamefont {Darqui{\'e}}, \citenamefont {Jones},
  \citenamefont {Jacubowiez}, \citenamefont {Browaeys},\ and\ \citenamefont
  {Grangier}}]{bergamini2004holographic}%
  \BibitemOpen
  \bibfield  {author} {\bibinfo {author} {\bibfnamefont {S.}~\bibnamefont
  {Bergamini}}, \bibinfo {author} {\bibfnamefont {B.}~\bibnamefont
  {Darqui{\'e}}}, \bibinfo {author} {\bibfnamefont {M.}~\bibnamefont {Jones}},
  \bibinfo {author} {\bibfnamefont {L.}~\bibnamefont {Jacubowiez}}, \bibinfo
  {author} {\bibfnamefont {A.}~\bibnamefont {Browaeys}}, \ and\ \bibinfo
  {author} {\bibfnamefont {P.}~\bibnamefont {Grangier}},\ }\href@noop {}
  {\bibfield  {journal} {\bibinfo  {journal} {JOSA B}\ }\textbf {\bibinfo
  {volume} {21}},\ \bibinfo {pages} {1889} (\bibinfo {year}
  {2004})}\BibitemShut {NoStop}%
\bibitem [{\citenamefont {Boyer}\ \emph {et~al.}(2006)\citenamefont {Boyer},
  \citenamefont {Godun}, \citenamefont {Smirne}, \citenamefont {Cassettari},
  \citenamefont {Chandrashekar}, \citenamefont {Deb}, \citenamefont {Laczik},\
  and\ \citenamefont {Foot}}]{boyer2006dynamic}%
  \BibitemOpen
  \bibfield  {author} {\bibinfo {author} {\bibfnamefont {V.}~\bibnamefont
  {Boyer}}, \bibinfo {author} {\bibfnamefont {R.}~\bibnamefont {Godun}},
  \bibinfo {author} {\bibfnamefont {G.}~\bibnamefont {Smirne}}, \bibinfo
  {author} {\bibfnamefont {D.}~\bibnamefont {Cassettari}}, \bibinfo {author}
  {\bibfnamefont {C.}~\bibnamefont {Chandrashekar}}, \bibinfo {author}
  {\bibfnamefont {A.}~\bibnamefont {Deb}}, \bibinfo {author} {\bibfnamefont
  {Z.}~\bibnamefont {Laczik}}, \ and\ \bibinfo {author} {\bibfnamefont
  {C.}~\bibnamefont {Foot}},\ }\href@noop {} {\bibfield  {journal} {\bibinfo
  {journal} {Phys. Rev. A}\ }\textbf {\bibinfo {volume} {73}},\ \bibinfo
  {pages} {031402} (\bibinfo {year} {2006})}\BibitemShut {NoStop}%
\bibitem [{\citenamefont {Pasienski}\ and\ \citenamefont
  {DeMarco}(2008)}]{pasienski2008high}%
  \BibitemOpen
  \bibfield  {author} {\bibinfo {author} {\bibfnamefont {M.}~\bibnamefont
  {Pasienski}}\ and\ \bibinfo {author} {\bibfnamefont {B.}~\bibnamefont
  {DeMarco}},\ }\href@noop {} {\bibfield  {journal} {\bibinfo  {journal} {Opt.
  Express}\ }\textbf {\bibinfo {volume} {16}},\ \bibinfo {pages} {2176}
  (\bibinfo {year} {2008})}\BibitemShut {NoStop}%
\bibitem [{\citenamefont {Gaunt}\ and\ \citenamefont
  {Hadzibabic}(2012)}]{gaunt2012robust}%
  \BibitemOpen
  \bibfield  {author} {\bibinfo {author} {\bibfnamefont {A.~L.}\ \bibnamefont
  {Gaunt}}\ and\ \bibinfo {author} {\bibfnamefont {Z.}~\bibnamefont
  {Hadzibabic}},\ }\href@noop {} {\bibfield  {journal} {\bibinfo  {journal}
  {Sci. Rep.}\ }\textbf {\bibinfo {volume} {2}},\ \bibinfo {pages} {1}
  (\bibinfo {year} {2012})}\BibitemShut {NoStop}%
\bibitem [{\citenamefont {Bender}\ \emph {et~al.}(1989)\citenamefont {Bender},
  \citenamefont {Milton}, \citenamefont {Pinsky},\ and\ \citenamefont
  {Simmons~Jr}}]{bender1989new}%
  \BibitemOpen
  \bibfield  {author} {\bibinfo {author} {\bibfnamefont {C.~M.}\ \bibnamefont
  {Bender}}, \bibinfo {author} {\bibfnamefont {K.~A.}\ \bibnamefont {Milton}},
  \bibinfo {author} {\bibfnamefont {S.~S.}\ \bibnamefont {Pinsky}}, \ and\
  \bibinfo {author} {\bibfnamefont {L.}~\bibnamefont {Simmons~Jr}},\
  }\href@noop {} {\bibfield  {journal} {\bibinfo  {journal} {J. Math. Phys.}\
  }\textbf {\bibinfo {volume} {30}},\ \bibinfo {pages} {1447} (\bibinfo {year}
  {1989})}\BibitemShut {NoStop}%
\bibitem [{\citenamefont {Hamam}\ \emph {et~al.}(2007)\citenamefont {Hamam},
  \citenamefont {Karalis}, \citenamefont {Joannopoulos},\ and\ \citenamefont
  {Solja{\v c}i{\'c}}}]{Hamam.2007.053801}%
  \BibitemOpen
  \bibfield  {author} {\bibinfo {author} {\bibfnamefont {R.~E.}\ \bibnamefont
  {Hamam}}, \bibinfo {author} {\bibfnamefont {A.}~\bibnamefont {Karalis}},
  \bibinfo {author} {\bibfnamefont {J.~D.}\ \bibnamefont {Joannopoulos}}, \
  and\ \bibinfo {author} {\bibfnamefont {M.}~\bibnamefont {Solja{\v
  c}i{\'c}}},\ }\href@noop {} {\bibfield  {journal} {\bibinfo  {journal} {Phys.
  Rev. A}\ }\textbf {\bibinfo {volume} {75}},\ \bibinfo {pages} {053801}
  (\bibinfo {year} {2007})}\BibitemShut {NoStop}%
\bibitem [{\citenamefont {Joannopoulos}\ \emph {et~al.}(2008)\citenamefont
  {Joannopoulos}, \citenamefont {Johnson}, \citenamefont {Winn},\ and\
  \citenamefont {Meade}}]{Joannopoulos.2008.Photonic}%
  \BibitemOpen
  \bibfield  {author} {\bibinfo {author} {\bibfnamefont {J.~D.}\ \bibnamefont
  {Joannopoulos}}, \bibinfo {author} {\bibfnamefont {S.~G.}\ \bibnamefont
  {Johnson}}, \bibinfo {author} {\bibfnamefont {J.~N.}\ \bibnamefont {Winn}}, \
  and\ \bibinfo {author} {\bibfnamefont {R.~D.}\ \bibnamefont {Meade}},\
  }\href@noop {} {\emph {\bibinfo {title} {Photonic Crystals: Molding the Flow
  of Light}}},\ \bibinfo {edition} {2nd}\ ed.\ (\bibinfo  {publisher}
  {Princeton University},\ \bibinfo {year} {2008})\BibitemShut {NoStop}%
\bibitem [{\citenamefont {Fan}\ \emph {et~al.}(2003)\citenamefont {Fan},
  \citenamefont {Suh},\ and\ \citenamefont {Joannopoulos}}]{Fan.2003.569}%
  \BibitemOpen
  \bibfield  {author} {\bibinfo {author} {\bibfnamefont {S.}~\bibnamefont
  {Fan}}, \bibinfo {author} {\bibfnamefont {W.}~\bibnamefont {Suh}}, \ and\
  \bibinfo {author} {\bibfnamefont {J.~D.}\ \bibnamefont {Joannopoulos}},\
  }\href@noop {} {\bibfield  {journal} {\bibinfo  {journal} {JOSA A}\ }\textbf
  {\bibinfo {volume} {20}},\ \bibinfo {pages} {569} (\bibinfo {year}
  {2003})}\BibitemShut {NoStop}%
\bibitem [{\citenamefont {Suh}\ \emph {et~al.}(2004)\citenamefont {Suh},
  \citenamefont {Wang},\ and\ \citenamefont {Fan}}]{Suh.2004.1511}%
  \BibitemOpen
  \bibfield  {author} {\bibinfo {author} {\bibfnamefont {W.}~\bibnamefont
  {Suh}}, \bibinfo {author} {\bibfnamefont {Z.}~\bibnamefont {Wang}}, \ and\
  \bibinfo {author} {\bibfnamefont {S.}~\bibnamefont {Fan}},\ }\href@noop {}
  {\bibfield  {journal} {\bibinfo  {journal} {IEEE J. Quantum Electron.}\
  }\textbf {\bibinfo {volume} {40}},\ \bibinfo {pages} {1511} (\bibinfo {year}
  {2004})}\BibitemShut {NoStop}%
\bibitem [{\citenamefont {Lane}\ and\ \citenamefont
  {Thomas}(1958)}]{Lane.1958.257}%
  \BibitemOpen
  \bibfield  {author} {\bibinfo {author} {\bibfnamefont {A.~M.}\ \bibnamefont
  {Lane}}\ and\ \bibinfo {author} {\bibfnamefont {R.~G.}\ \bibnamefont
  {Thomas}},\ }\href@noop {} {\bibfield  {journal} {\bibinfo  {journal} {Rev.
  Mod. Phys.}\ }\textbf {\bibinfo {volume} {30}},\ \bibinfo {pages} {257}
  (\bibinfo {year} {1958})}\BibitemShut {NoStop}%
\bibitem [{\citenamefont {Leal}(1999)}]{Leal.2010.Neutron}%
  \BibitemOpen
  \bibfield  {author} {\bibinfo {author} {\bibfnamefont {L.~C.}\ \bibnamefont
  {Leal}},\ }\href@noop {} {\bibfield  {journal} {\bibinfo  {journal} {Brief
  review of the {R}-Matrix theory, SAMMY}\ } (\bibinfo {year}
  {1999})}\BibitemShut {NoStop}%
\bibitem [{\citenamefont {Hager}(1989)}]{Hager_1989}%
  \BibitemOpen
  \bibfield  {author} {\bibinfo {author} {\bibfnamefont {W.~W.}\ \bibnamefont
  {Hager}},\ }\href@noop {} {\bibfield  {journal} {\bibinfo  {journal} {SIAM
  Rev.}\ }\textbf {\bibinfo {volume} {31}},\ \bibinfo {pages} {221} (\bibinfo
  {year} {1989})}\BibitemShut {NoStop}%
\end{thebibliography}%

\appendix

\section{Quantum RSM Theory\label{sec:RSMforQuantum}}

To emphasize the applicability of RSM theory to quantum mechanics, we prove that an infinite number of R-zeros exist in suitable  quantum-scattering systems. This adapts the temporal coupled-mode theory proof in optics \cite{Sweeney.2020.063511,Stone.2020.343} to quantum mechanics via the Breit-Wigner approximation to multichannel quantum $R$-matrix theory and generalizes the recently identified  phenomenological link between temporal coupled-mode theory and Breit-Wigner cross sections \cite{Hamam.2007.053801,Joannopoulos.2008.Photonic}
to the nonoverlapping-resonance case.

\subsection{Relationship between temporal coupled-mode theory and quantum R-matrix theory}

We begin by developing an expression for the Breit-Wigner inelastic cross section
that assumes the form of the temporal coupled-mode theory formula
\begin{align}
\mathbf{S}\left(\omega\right)&=\Big(\mathbf{I}_{N}-\text{i}\mathbf{D}\frac{1}{\omega-\mathbf{H}_{\text{eff}}}\mathbf{D}^{\dagger}\Big)\mathbf{S}_{0},\nonumber\\
\mathbf{H}_{\text{eff}}&=\mathbf{H}_{\text{close}}-\textstyle{\frac{\text{i}}{2}}\mathbf{D}^{\dagger}\mathbf{D},\nonumber
\end{align}
where $\mathbf{S}(\omega)$ is the scattering matrix as a function of the frequency $\omega$, $\mathbf{D}$ is the coupling matrix, $\mathbf{S}_0$ is the background scattering matrix, $\mathbf{H}_\text{eff}$ is the effective Hamiltonian, and $\mathbf{H}_\text{close}$ is the Hamiltonian associated with the closed scattering region (the resonances) \cite{Fan.2003.569,Suh.2004.1511}. 

The initial expression is given by the Breit-Wigner approximation to the inelastic cross section in level-matrix-format multichannel quantum $R$-matrix theory \cite{Lane.1958.257,Leal.2010.Neutron}
\begin{align}
U_{cc^{\prime}}(E)  &=\Bigg[\delta_{cc^{\prime}}-\text{i}\sum_{\lambda\mu}\Gamma_{\lambda c}^{1/2}\frac{1}{E-\left(E_{\lambda}+\textstyle{\frac{\text{i}}{2}}\Gamma_{\lambda}\right)}\Gamma_{\mu c^{\prime}}^{1/2}\Bigg]\nonumber\\
&\times\text{e}^{-\text{i}\left(\phi_{c}+\phi_{c^{\prime}}\right)},\nonumber
\end{align}
where $U_{cc^{\prime}}$ is the collision matrix element that connects the incoming $c$ and outgoing $c^\prime$ channels as a function of the energy $E$, $\delta_{cc^{\prime}}$ is the Kronecker delta, $\phi_c$ and $\phi_{c^{\prime}}$ define the relative phase of the channels, and $E_\lambda$ is the eigenenergy of the resonance wavefunctions $X_\lambda$. The coupling between the the resonance wavefunctions   $X_\lambda$ and the channel wavefunctions $\varphi_c^{\star}$ at a dividing surface $S$ is given by the reduced-width amplitude $\gamma_{\lambda c} =\sqrt{{\hbar^{2}}/({2M_{c}a_{c}}})\int\text{d}S\varphi_{c}^{\star}X_{\lambda}$ for system mass $M_c$ and channel radius $a_c$.  The coupling matrix elements $\Gamma_{\lambda}=\sum_{c}\Gamma_{\lambda c}$ are given by $\Gamma_{\mu c}^{1/2} =\gamma_{\mu c}\left(2P_{c}\right)^{1/2}$ for the incoming wave at the channel radius $I_c$, the outgoing wave at channel radius $O_c$, and the derived term $\mathbf{P}=\left(\mathbf{I}\mathbf{O}\right)^{-1}$.

We proceed by recognizing that the scattering matrix is equivalent to the collision matrix $\mathbf{S}=\mathbf{U}$ and the matrix form of the delta function is the identity
matrix $\mathbf{I}_{N,cc^{\prime}}\equiv\delta_{cc^{\prime}}$.
We then define the background scattering in terms of the relative phase of the channels $\mathbf{S}_{0,cc^{\prime}}\equiv\text{e}^{-\text{i}\left(\phi_{c}+\phi_{c^{\prime}}\right)}$.
Together, this yields
\begin{align}
\mathbf{S}_{cc^{\prime}} & =\bigg[\mathbf{I}_{cc^{\prime}}-\text{i}\sum_{\lambda\mu}\Gamma_{\lambda c}^{1/2}\frac{1}{E-\left(E_{\lambda}+\textstyle{\frac{\text{i}}{2}}\Gamma_{\lambda}\right)}\Gamma_{\mu c^{\prime}}^{1/2}\bigg]\mathbf{S}_{0,cc^{\prime}}.\nonumber
\end{align}

To reexpress the fractional term, we define a column coupling vector in terms of the coupling matrix elements $\mathbf{d}_{\lambda c}\equiv\Gamma_{\lambda c}^{1/2}$,
where the coupling to the input channel $c$ is the complex conjugate
of the coupling to the output channel $c^{\prime}$. The overall coupling
to a given resonance is then $\mathbf{d}_{\lambda}^{\dagger}\mathbf{d}_{\lambda} =\sum_{c}\lvert \Gamma_{\lambda c}^{1/2}\rvert ^{2}
=\sum_{c}\Gamma_{\lambda c} =\Gamma_{\lambda}$,
such that the denominator of the fractional term in the Breit-Wigner cross section is
\begin{equation}
E-\big(E_{\lambda}+\textstyle{\frac{\text{i}}{2}}\Gamma_{\lambda}\big)=E-\big(E_{\lambda}+\textstyle{\frac{\text{i}}{2}}\mathbf{d}_{\lambda}^{\dagger}\mathbf{d}_{\lambda}\big).\nonumber
\end{equation}
Te numerator is independent of $\mu$, so the fractional
term in the Breit-Wigner cross section is
\begin{align}
\sum_{\lambda\mu}&\Gamma_{\lambda c}^{1/2}\frac{1}{E-\left(E_{\lambda}+\textstyle{\frac{\text{i}}{2}}\Gamma_{\lambda}\right)}\Gamma_{\mu c^{\prime}}^{1/2}\nonumber \\
 & =\sum_{\lambda}\Gamma_{\lambda c}^{1/2}\frac{1}{E-\left(E_{\lambda}+\textstyle{\frac{\text{i}}{2}}\mathbf{d}_{\lambda}^{\dagger}\mathbf{d}_{\lambda}\right)}\sum_{\mu}\Gamma_{\mu c^{\prime}}^{1/2}\nonumber\\
 & =\bigg[\sum_{\lambda}\mathbf{d}_{\lambda}\frac{1}{E-\left(E_{\lambda}+\textstyle{\frac{\text{i}}{2}}\mathbf{d}_{\lambda}^{\dagger}\mathbf{d}_{\lambda}\right)}\sum_{\mu}\mathbf{d}_{\mu}^{\dagger}\bigg]_{cc^{\prime}}.\nonumber
\end{align}
Let the coupling matrix be defined by the coupling vectors as $\mathbf{D}=\big[\begin{array}{cccc}
\mathbf{d}_{\lambda_{1}} & \mathbf{d}_{\lambda_{2}} & \cdots & \mathbf{d}_{\lambda_{n}}\end{array}\big]^{T}$.
This gives the result
\begin{equation}
\begin{gathered}
\sum_{\lambda}\mathbf{d}_{\lambda}\frac{1}{E-\big(E_{\lambda}+\textstyle{\frac{\text{i}}{2}}\mathbf{d}_{\lambda}^{\dagger}\mathbf{d}_{\lambda}\big)}\sum_{\mu}\mathbf{d}_{\mu}^{\dagger}\nonumber\\
=\mathbf{D}\frac{1}{E-\big(E_{\lambda}+\textstyle{\frac{\text{i}}{2}}\mathbf{D}^{\dagger}\mathbf{D}\big)}\mathbf{D}^{\dagger}\nonumber
\end{gathered}
\end{equation}
and the scattering matrix
\begin{align}
\mathbf{S}(E) & =\bigg[\mathbf{I}-\text{i}\mathbf{D}\frac{1}{E-\left(\mathbf{E}_{\lambda}+\textstyle{\frac{\text{i}}{2}}\mathbf{D}^{\dagger}\mathbf{D}\right)}\mathbf{D}^{\dagger}\bigg]\mathbf{S}_{0},\nonumber
\end{align}
where $\mathbf{E}_{\lambda}$ is now an $M\times M$ diagonal matrix (the same shape as $\mathbf{D}^{\dagger}\mathbf{D}$) of the $M$
(theoretically infinite in number) resonance energies. 

Finally, to express the Breit-Wigner cross section in temporal coupled-mode theory form, we identify the system Hamiltonian.
The closed Hamiltonian is the resonance Hamiltonian $\mathbf{H}_{\text{close}}=\mathbf{E}_{\lambda}$
as the resonances completely describe the behavior of the system within the aforementioned surface $S$. The effective Hamiltonian is then the sum of the resonance Hamiltonian and its coupling to external channels $\mathbf{H}_\text{eff}=\mathbf{E}_{\lambda}+\textstyle{\frac{\text{i}}{2}}\mathbf{D}^{\dagger}\mathbf{D}$.
This yields the temporal coupled-mode  form of the Breit-Wigner cross section
\begin{align}
\mathbf{S}\left(E\right) & =\bigg(\mathbf{I}_{N}-\text{i}\mathbf{D}\frac{1}{E-\mathbf{H}_{\text{eff}}}\mathbf{D}^{\dagger}\bigg)\mathbf{S}_{0}.\nonumber
\end{align}

\subsection{Proof of an infinite number of R-zero modes}

To prove that there are an infinite number R-zeros defined as the energies $E$, where the reflection matrix $\mathbf{R}_\text{in}(E)$ is zero, we find the energies at which the {\em inverse} reflection matrix $\mathbf{R}_\text{in}(E)$ is {\em infinite}. Let $\mathbf{F}$ be a filtering matrix  that eliminates all but the incoming
channels and $\mathbf{\bar{F}}$ be a filtering matrix that eliminates all but the outgoing channels. By definition, the filtering matrix and its complex conjugate yield $\mathbf{F}\mathbf{F}^{\dagger}=\mathbf{I}_{N_\text{in}}$,
 which reduces the dimension
of the channel space from $N$ to $N_\text{in}$ (that is $F_{ij}=\delta_{ij}$ for $i\leq N_\text{in}$, $j\leq N_\text{in}$). Also, by definition $\mathbf{F}\mathbf{F}^\dagger=I_{N_\text{out}}$.
The reflection matrix is then
\begin{align}
\mathbf{R}_{\text{in}}\left(E\right) & =\mathbf{F}\mathbf{S}\left(E\right)\mathbf{F}^{\dagger} =\mathbf{F}\bigg(\mathbf{I}_{N}-\text{i}\mathbf{D}\frac{1}{E-\mathbf{H}_{\text{eff}}}\mathbf{D}^{\dagger}\bigg)\mathbf{S}_{0}\mathbf{F}^{\dagger}.\nonumber
\end{align}
Repeated application of the filtering matrix yields no change because a projection
onto a projection yields the same matrix), so we are free to employ a second
round of filtering matrices
\begin{align}
\mathbf{R}_{\text{in}}\left(E\right)
 & =\Big(\mathbf{F}\Big[\mathbf{I}_{N}-\text{i}\mathbf{D}\frac{1}{E-\mathbf{H}_{\text{eff}}}\mathbf{D}^{\dagger}\Big]\mathbf{F}^{\dagger}\Big)\left(\mathbf{F}\mathbf{S}_{0}\mathbf{F}^{\dagger}\right)\nonumber\\
&=\Big(\mathbf{I}_{N_{\text{in}}}-\text{i}\mathbf{F}\mathbf{D}\frac{1}{E-\mathbf{H}_{\text{eff}}}\mathbf{D}^{\dagger}\mathbf{F}^{\dagger}\Big)\mathbf{S}_{0,\text{in}}.\nonumber
\end{align}

The inverse is then determined according to the equation for the inverse
of a matrix product $\left(\mathbf{AB}\right)^{-1}=\mathbf{B}^{-1}\mathbf{A}^{-1}$,
which gives
\begin{align}
\mathbf{R}_{\text{in}}^{-1}\left(E\right) & =\mathbf{S}_{0,\text{in}}^{-1}\Big(\mathbf{I}_{N_{\text{in}}}-\text{i}\mathbf{F}\mathbf{D}\frac{1}{E-\mathbf{H}_{\text{eff}}}\mathbf{D}^{\dagger}\mathbf{F}^{\dagger}\Big)^{-1}.\nonumber
\end{align}
The second term is evaluated via the Woodbury matrix identity \cite{Hager_1989}
\begin{equation}
\begin{gathered}
\left(\mathbf{A}+\mathbf{UCV}\right)^{-1}\nonumber\\
=\mathbf{A}^{-1}-\mathbf{A}^{-1}\mathbf{U}\left(\mathbf{C}^{-1}+\mathbf{V}\mathbf{A}^{-1}\mathbf{U}\right)^{-1}\mathbf{V}\mathbf{A}^{-1},\nonumber
\end{gathered}
\end{equation}
where
$\mathbf{A} =\mathbf{I}_{N_{\text{in}}}$,
$\mathbf{U}  =\mathbf{FD}$,
$\mathbf{C}  =-\text{i}/\left(E-\mathbf{H}_{\text{eff}}\right)$, and
$\mathbf{V}  =\mathbf{D}^{\dagger}\mathbf{F}^{\dagger}$. The above results give 
\begin{align}
\mathbf{R}_{\text{in}}^{-1}\left(E\right) & =\mathbf{S}_{0,\text{in}}^{-1}\Bigg(\mathbf{I}_{N_{\text{in}}} -\left(\mathbf{FD}\right)\Bigg.\nonumber\\
&\times\Bigg.\Big[\Big(-\frac{\text{i}}{E-\mathbf{H}_{\text{eff}}}\Big)^{-1}+\text{i}\mathbf{D}^{\dagger}\mathbf{F}^{\dagger}\mathbf{FD}\Big]^{-1}\left(\mathbf{D}^{\dagger}\mathbf{F}^{\dagger}\right)\Bigg)\nonumber\\
&=\mathbf{S}_{0,\text{in}}^{-1}\Bigg(\mathbf{I}_{N_{\text{in}}}+\text{i}\left(\mathbf{FD}\right)\Bigg.\nonumber\\
&\times\Bigg.\frac{1}{E-\mathbf{H}_{\text{eff}}-\text{i}\mathbf{D}^{\dagger}\mathbf{F}^{\dagger}\mathbf{FD}}\left(\mathbf{D}^{\dagger}\mathbf{F}^{\dagger}\right)\Bigg).\nonumber
\end{align}

To determine the coupling matrices in reduced-channel form, we define the filtered coupling matrices $\mathbf{D}_{\text{in}}\equiv\mathbf{F}\mathbf{D}$
and $\mathbf{D}_{\text{out}}\equiv\bar{\mathbf{F}}\mathbf{D}$
and express the effective Hamiltonian $\mathbf{H}_\text{eff}$ in terms of its components, which yields
\begin{gather}
\mathbf{R}_{\text{in}}^{-1}\left(E\right) =\mathbf{S}_{0,\text{in}}^{-1}\nonumber\\
\times\left(\mathbf{I}_{N_{\text{in}}}+\text{i}\mathbf{D}_{\text{in}}\frac{1}{E-\mathbf{H}_{\text{close}}+\textstyle{\frac{\text{i}}{2}}\mathbf{D}^{\dagger}\mathbf{D}-\text{i}\mathbf{D}^{\dagger}\mathbf{F}^{\dagger}\mathbf{FD}}\mathbf{D}_{\text{in}}^{\dagger}\right).\nonumber
\end{gather}
Subsequent application of the filtering matrix identity
$\mathbf{F}^{\dagger}\mathbf{F}+\bar{\mathbf{F}}^{\dagger}\bar{\mathbf{F}}=\mathbf{I}_{N}$
gives the complete inverse reflection matrix
\begin{align}
\mathbf{R}_{\text{in}}^{-1}\left(E\right) & =\mathbf{S}_{0,\text{in}}^{-1}\Big(\mathbf{I}_{N,\text{in}}+\text{i}\mathbf{D}_{\text{in}}\frac{1}{E-\mathbf{H}_{\text{eff}}}\mathbf{D}_{\text{in}}^{\dagger}\Big),\nonumber\\
\mathbf{H}_{\text{eff}} & \equiv\mathbf{H}_{\text{close}}+\textstyle{\frac{\text{i}}{2}}\mathbf{D}_{\text{in}}^{\dagger}\mathbf{D}_{\text{in}}-\textstyle{\frac{\text{i}}{2}}\mathbf{D}_{\text{out}}^{\dagger}\mathbf{D}_{\text{out}},\nonumber
\end{align}
in direct analogy to temporal coupled-mode theory. 

We now consider the number of R-zeros entailed by the expression. Since the background scattering $\mathbf{S}_0$ amounts to a phase factor, $\mathbf{R}_{\text{in}}^{-1}$ has no poles and does not play a role in the existence of R-zeros. The RSMs are therefore wholly determined by poles of the
effective Hamiltonian $\mathbf{H}_\text{eff}$. When there are an infinite number of resonances $E_\lambda$, there are an infinite number of eigenvalues of the resonance  $\mathbf{H}_\text{close}$ and effective $\mathbf{H}_\text{eff}$ Hamiltonians. Thus, there are an infinite number of solutions for which $E-\mathbf{H}_\text{eff}=0$. In this case, the inverse reflection matrix is singular, so there are an infinite number of singularities in the inverse reflection matrix $\mathbf{R}^{-1}_\text{in}(E)$ and an infinite number of R-zeros in the reflection matrix $\mathbf{R}_\text{in}(E)$.

\subsection{Relationship between applications of RSM theory to optics and quantum mechanics}

RSM theory in quantum mechanics is fundamentally different from RSM theory in optics.  The susceptibility $\varepsilon(x)$ multiplies the eigenvalue in the Maxwell equation $\Big(\nabla^2+\frac{\epsilon(\bf{x})\omega^2}{c^2}\Big)\vec{E}(\bf{x})=0$, but the potential $V(x)$ in the Schr{\"o}dinger equation  $\Big(-\frac{\hbar^2}{2m}\nabla^2+V(x)-E\Big)\psi({\bf{x}})=0$
does not. This special property of quantum mechanics accounts for the unique above-barrier quantum reflection effect examined in this paper.

\section{Truncation Lengths and Energy Bounds\label{sec:BoundedEnergy}}

\begin{figure*}
\centering
\includegraphics[height=0.25\textheight]{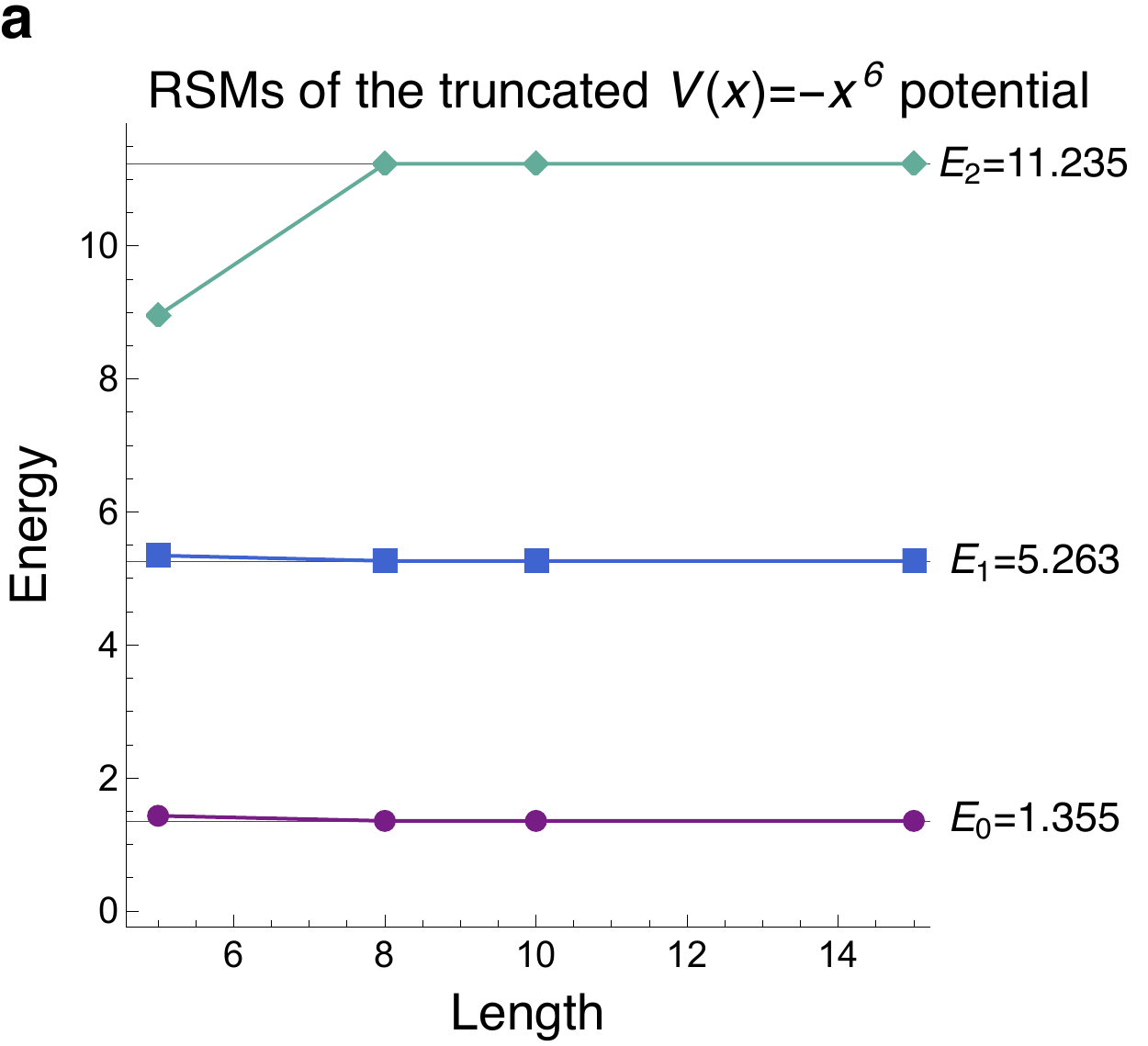}\includegraphics[height=0.25\textheight]{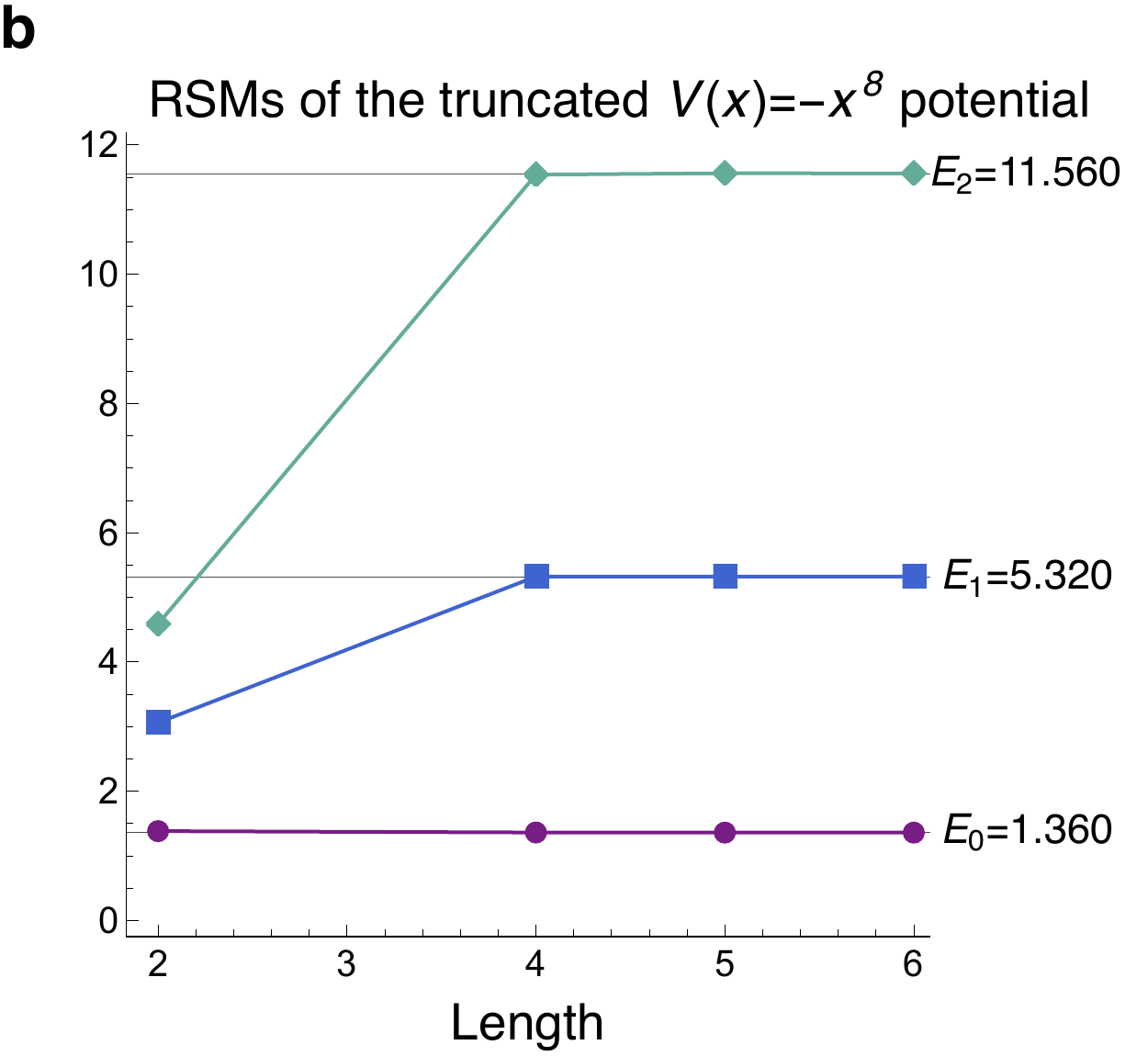}
\caption{RSMs of the truncated $V(x)=-\lvert x\rvert ^p$ potential for (a) $p=6$ and (b) $p=8$ (colored lines with points) agree with the  weakly-bound state energies $E_i$ for $i=0-2$ of the infinite-length potentials (gray horizontal lines) for longer truncation lengths $L$. \label{fig:RSMTruncatedResultsMore}}
\end{figure*}

\begin{figure*}
\centering
\includegraphics[width=1\textwidth]{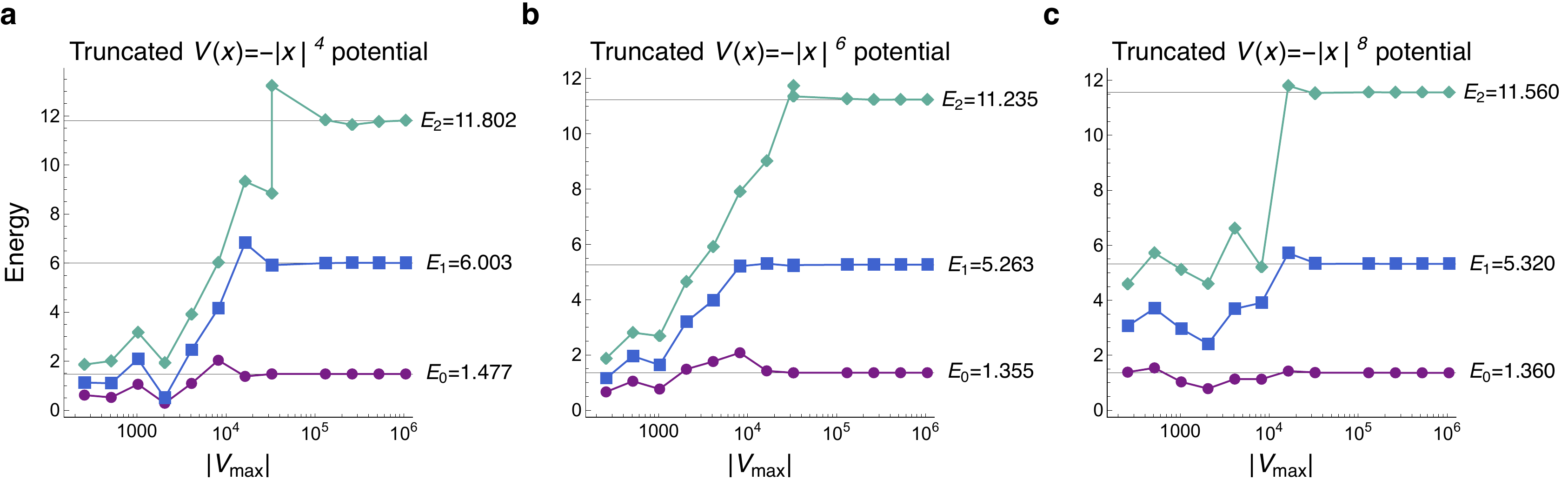}
\caption{RSMs of the truncated $V(x)=-\lvert x\rvert ^p$ potential (colored lines with points) are found to reproduce the analytic weakly-bound state energies $E_i$ for $i=0-2$ of the infinite-length (a) $V(x)=-x^4$, (b) $-x^6$, and (c) $-x^8$ potentials (gray horizontal lines) for sufficiently large energy bound $\lvert V_{\max}\rvert $. \label{fig:BoundedEnergy}}
\end{figure*}

As shown in Fig.~\ref{fig:RSMTruncatedResults}(a), Fig.~\ref{fig:RSMTruncatedResultsMore}, and Fig.~\ref{fig:BoundedEnergy}, the RSMs of the truncated $V(x)=-\lvert x\rvert ^p$ potential accurately reproduce the known analytic eigenenergies of the weakly-bound states of the $V(x)=-x^4$, $-x^6$, and $-x^8$ potentials for sufficiently large truncation length $L$ and energy bound $\lvert V_{\max}\rvert $. Similar to the RSMs as a function of truncation length, arbitrarily higher-lying eigenenergies are determined as the energy bound is increased, with the eigenenergies for $p=4$ most sensitive to changes in the energy bound.

\section{Quantum-Scattering Calculations\label{sec:QuantumScatteringTruncation}}

We support our RSM theory calculations with traditional quantum-scattering techniques following Ref.~\cite{Bender.2018.052118}. According to the exact quantum numerical approach, the scattering wavefunction is divided into three regions: (I)
$x<L$, (II) $-L\le x\le L$, and (III) $x>L$. In Region I the wavefunction
is expressed as a sum of incident and reflected waves at energy $E$ 
\begin{equation}
\psi(x)=\text{e}^{\text{i}\sqrt{E}x}+R\text{e}^{-\text{i}\sqrt{E}x},\nonumber
\end{equation}
and in Region III the wavefunction is expressed only in terms of
the amplitude-$T$ transmitted wave $T\exp\left(ix\sqrt{E}\right)$.
The scattering wavefunction is determined by integrating the time-dependent
Schr{\"o}dinger equation from the lower limit of region III $y_{\text{III}}\left(L\right)  =1, y_{\text{III}}^{\prime}\left(L\right)  =\text{i}\sqrt{E}$
to the upper limit of region I
\begin{align}
y_{\text{I}}\left(-L\right) & =e ^{-2\text{i}L\sqrt{E}}/T+R/T,\nonumber\\
y_{\text{I}}^{\prime}\left(-L\right) & =\text{i}\sqrt{E}e^{-2\text{i}L\sqrt{E}}/T-\text{i}R\sqrt{E}/T,\nonumber
\end{align}
where $y\left(x\right)$ is the scaled scattering wavefunction
\begin{equation}
y\left(x\right)\equiv\psi\left(x\right)\text{e}^{-\text{i}L\sqrt{E}}/T.\nonumber
\end{equation}
This expression allows the ratio of the reflection and transmission
coefficients to be calculated in terms of the wavefunction value at
$x=-L$
\begin{equation}
\lvert R/T\rvert =\textstyle{\frac{1}{2}}\lvert y\left(-L\right)+\text{i}y^{\prime}\left(-L\right)/\sqrt{E}\rvert .\nonumber
\end{equation}
Similarly, the reflection ratio of the left-moving wave 
\begin{equation}
    \psi=Re^{\text{i}\sqrt{E}x}+e^{-\text{i}\sqrt{E}x}\nonumber
\end{equation}
with boundary conditions $y_{I}\left(-L\right)=1, y_{I}^{\prime}\left(-L\right)=-\text{i}\sqrt{E}$
is
\begin{equation}
    \lvert R/T\rvert =\textstyle{\frac{1}{2}}\lvert y_{III}\left(L\right)-\text{i}y_{III}^{\prime}\left(L\right)/\sqrt{E}\rvert. \nonumber
\end{equation}

The reflectance follows as
\begin{equation}
    \lvert R\rvert ^2=\left(\frac{\lvert R/T\rvert }{\sqrt{1+\lvert R/T\rvert ^2}}\right)^2\nonumber.
\end{equation}

\begin{figure}
\centering
\includegraphics[width=1\columnwidth]{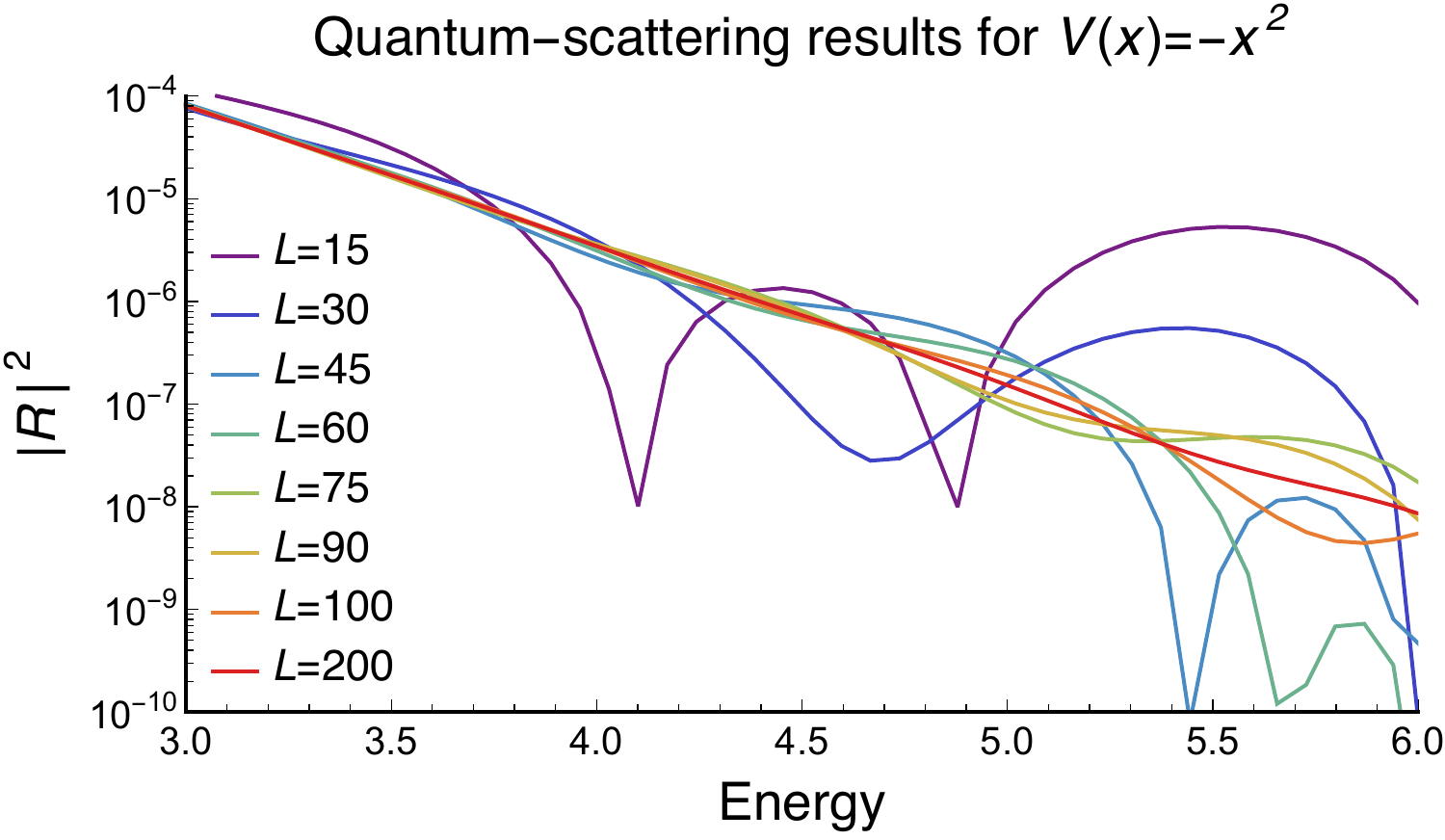}
\caption{Quantum-scattering theory results for the truncated $V(x)=-x^2$ reproduce the RSM theory results, as the system accurately reproduces the lack of real eigenenergies of the infinite-length potential within an arbitrarily wide energy domain as $L$ increases ({\em e.g.} no R-zeros or RSMs). For the truncated potential with $L=200$ (red line), no reflectionless scattering modes occur in the energy range $E\in[3.0,6.0]$, in agreement with the infinite-length result. \label{fig:QuantumScattering}}
\end{figure}

\begin{figure*}
\centering
\includegraphics[width=0.44\textwidth]{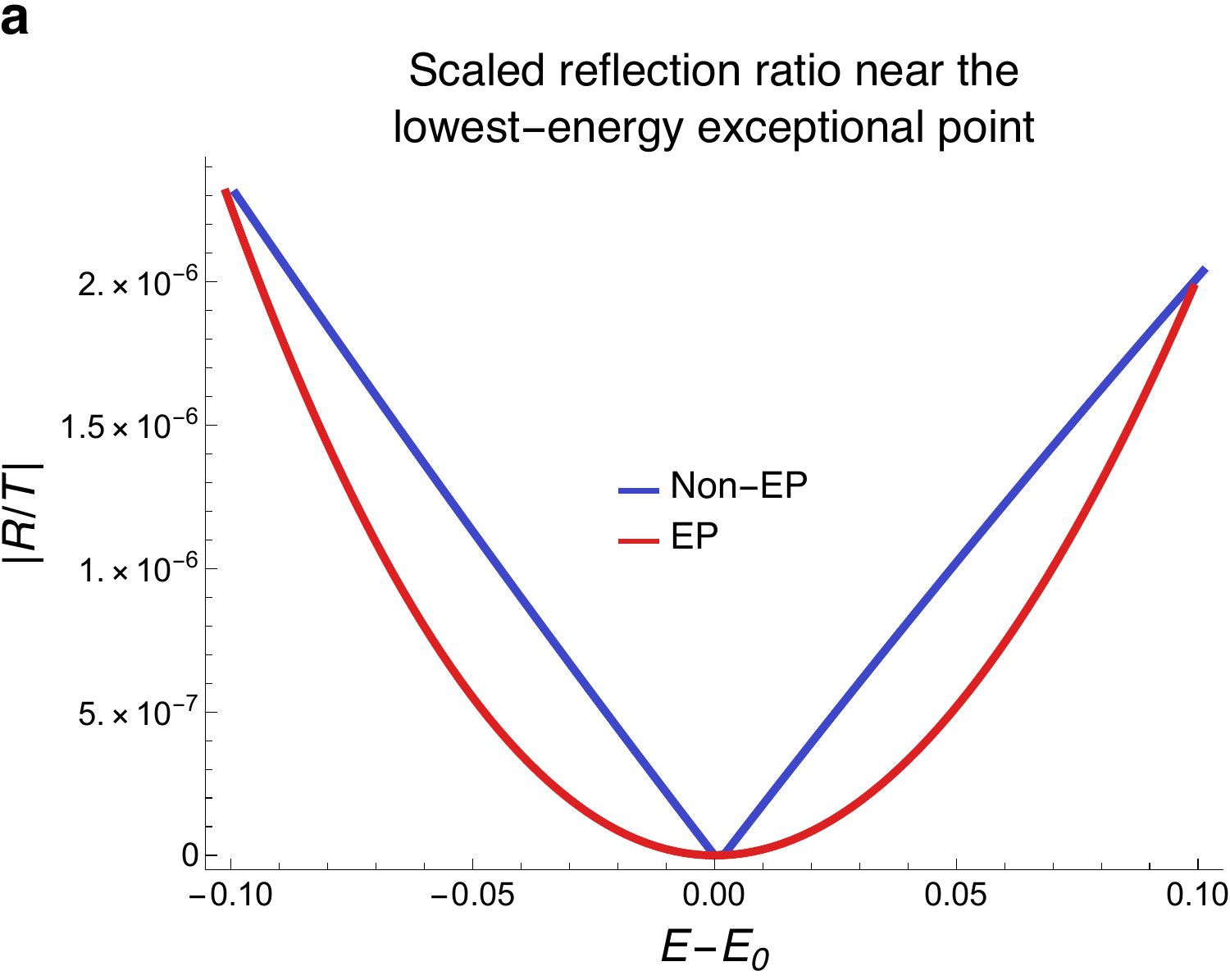}\qquad\includegraphics[width=0.44\textwidth]{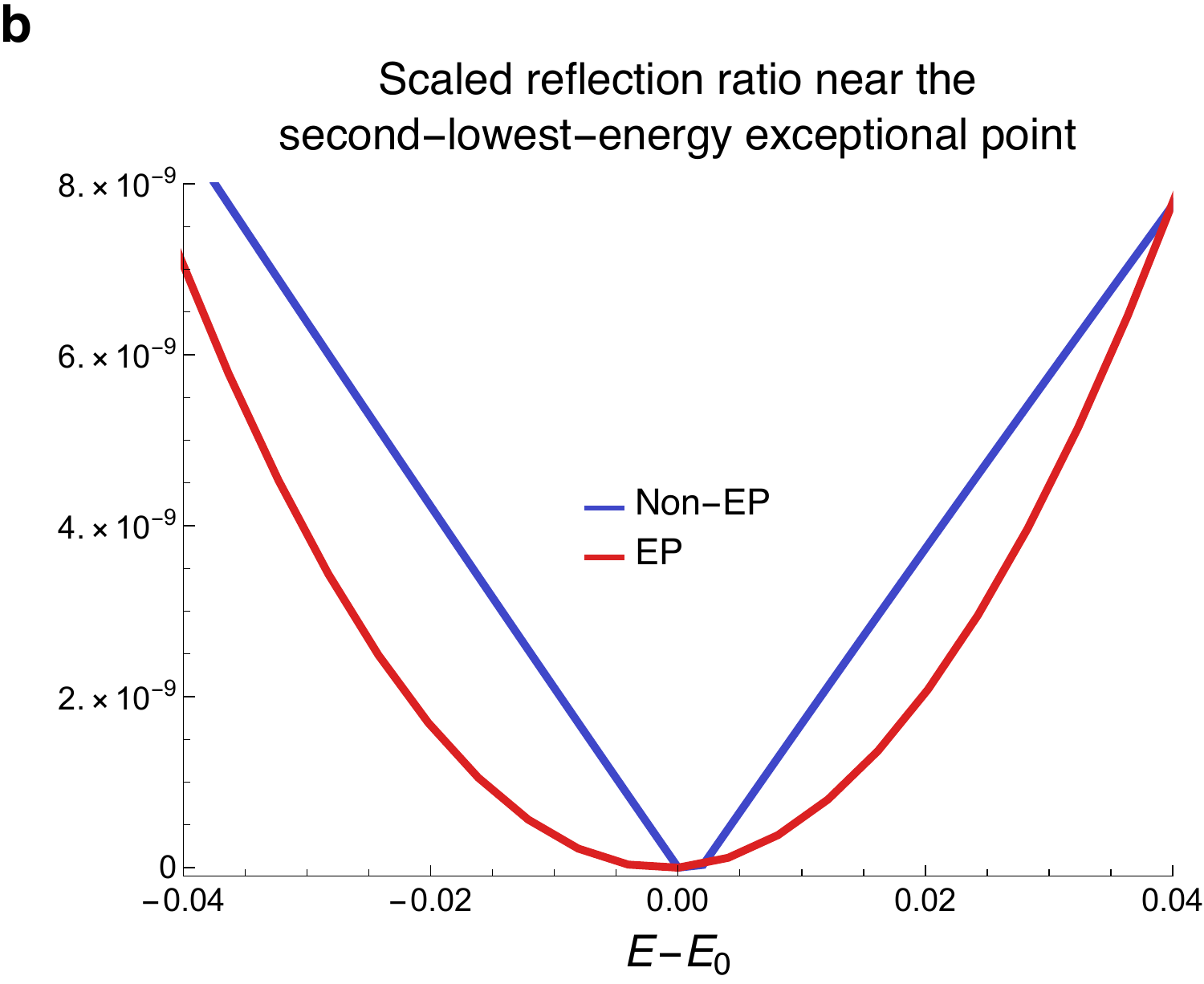}
\caption{Transitions from a linear (blue line) to quadratic reflectance ratio (red line) at the (a) lowest-energy and (b) second-lowest energy exceptional points. The nonexceptional point scaling is illustrated for the RSM near $p=4,E_1=6.00$. As expected, the truncation yields a shift in the measured RSMs (non-EP, $p=4,E_0=5.97$ for $L=15$; lowest-energy EP
$p=3.4,E_0=8.50$ for $L=50$; and second-lowest-energy EP $p=3.8,E_0=22.68$ for $L=50$).
\label{fig:EPmore}}
\end{figure*}

We consider the quantum-scattering results for the class of truncated $V(x)=-\lvert x\rvert ^p$ potentials.
As shown in Fig.~\ref{fig:QuantumScattering}, the quantum-scattering results for the $p=2$ potential indicate an absence of R-zeros in the energy range studied for sufficiently large $L$, in agreement with the results of RSM theory and the known monotonic decay of the reflection coefficient \cite{kemble1935contribution} and absence of weakly-bound states in the infinite-length $V(x)=-x^2$ potential \cite{Bender.2007.947}. 
As shown in Fig.~\ref{fig:EP}(b) and  Fig.~\ref{fig:EPmore}, the quantum-scattering calculations also demonstrate signatures of exceptional points, such as the transition from a linear reflectance ratio near the RSM to a quadratic reflectance ratio at the exceptional point (this is associated with enhanced sensitivity \cite{liu2016metrology}). 

\section{Extended WKB Force Potentials\label{sec:WKBTruncation}}

\begin{figure*}
\centering
\includegraphics[width=0.46\textwidth]{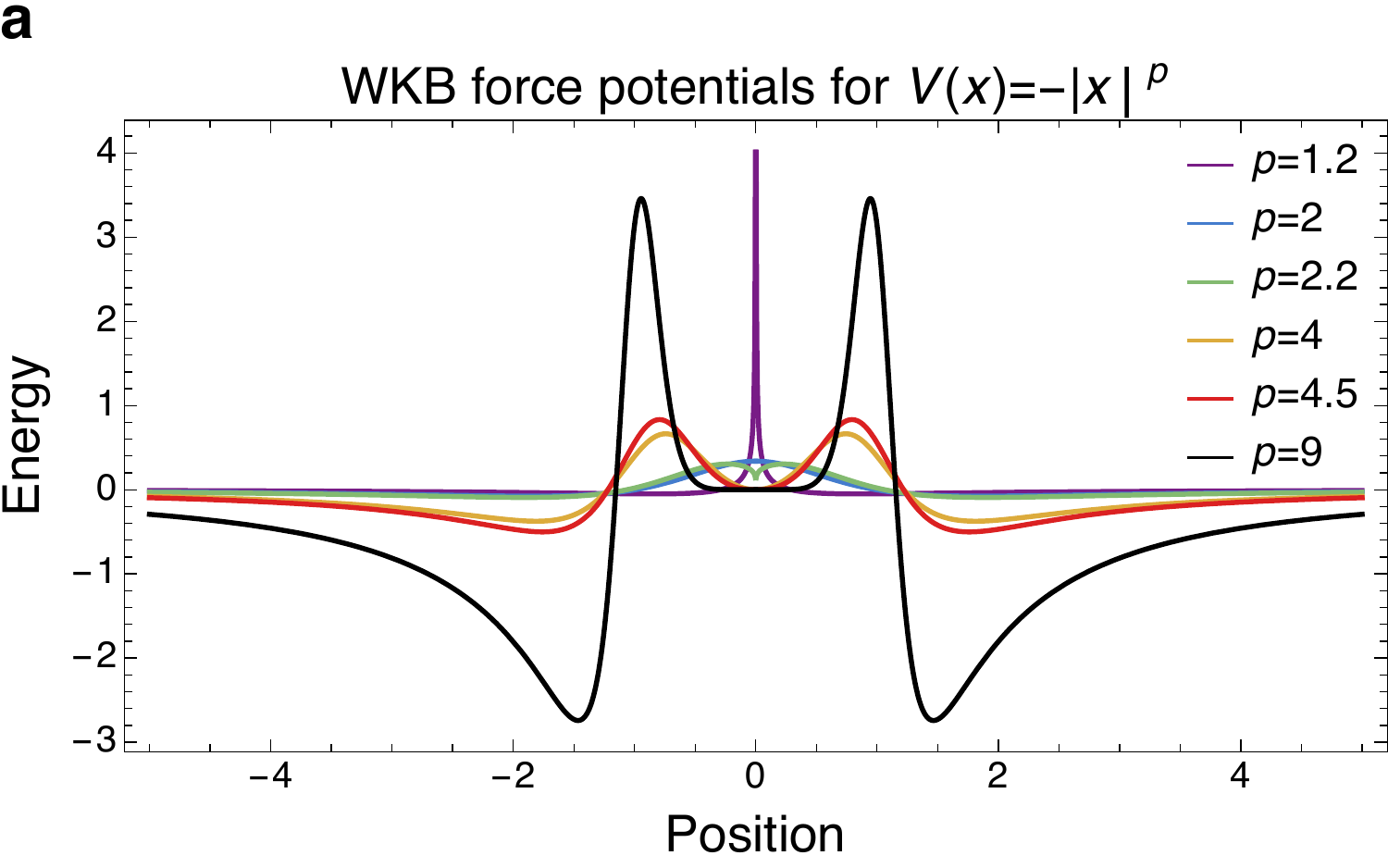}\quad\includegraphics[width=0.46\textwidth]{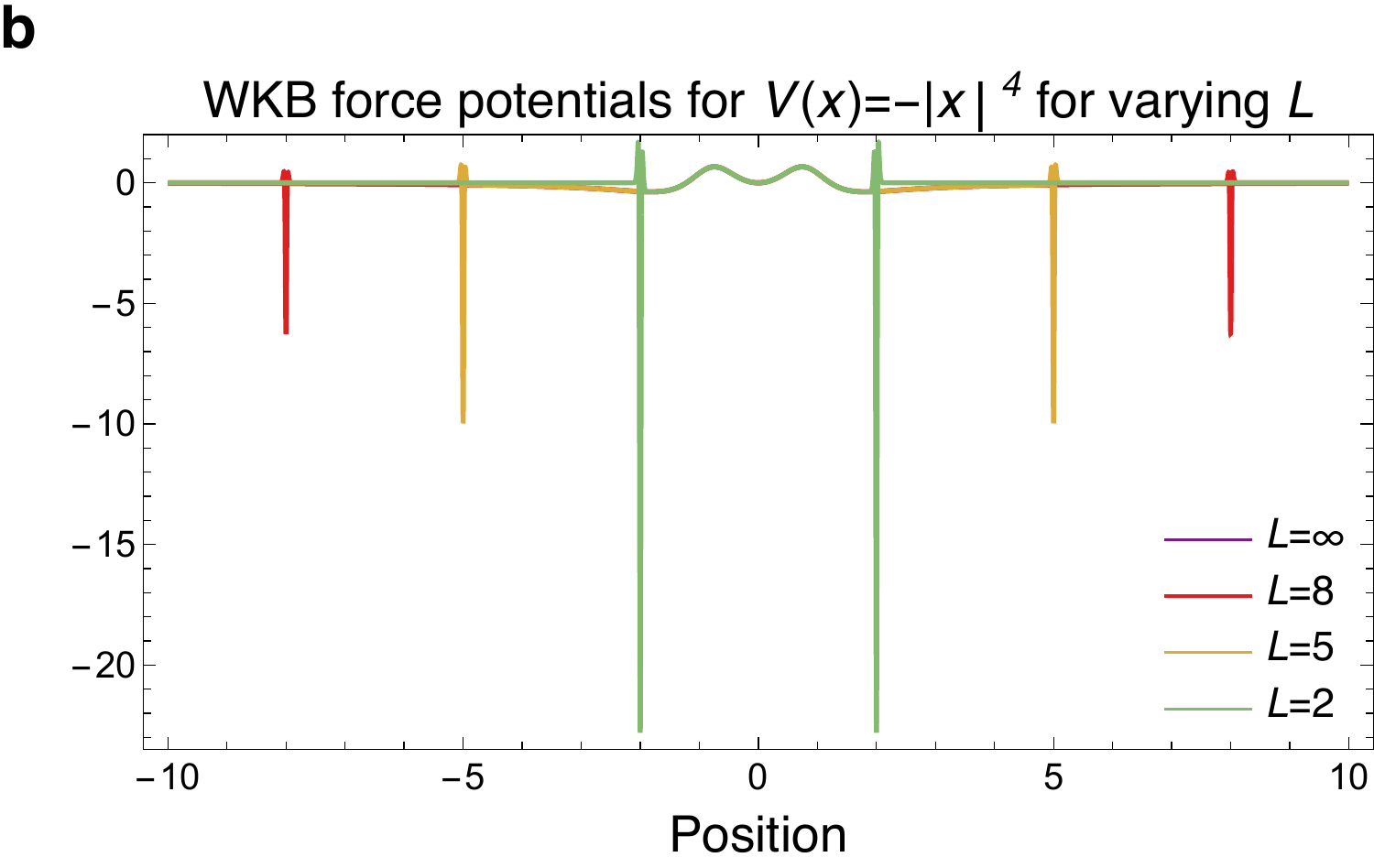}
\caption{WKB force potentials at $E=1.477$ for (a) the infinite-length $V(x)=-\lvert x\rvert ^p$ potential for varying $p$ and (b) the truncated $V(x)=-\lvert x\rvert ^4$ (colored lines), extending the magnitude and breadth shown in Fig.~\ref{fig:WKB}.   The height of the twin peaks about the origin in the $V(x)=-\lvert x\rvert ^p$ is seen to increase as $p$ increases beyond $p=4$, in agreement with the deeper dips in the reflection coefficient at the RSM energies at higher $p$  values. For the $L$ values depicted, the peaks are found to be largest for $L=2$ (green medium-long dashed line), followed by $L=5$ (yellow medium dashed line) and $L=8$ (red fine dashed line), in agreement with the prediction that anomalous quantum reflection due to truncation decreases as $L$ increases. \label{fig:WKBTruncation}}
\end{figure*}

The behavior of the WKB force potential for the $V(x)=-\lvert x\rvert ^p$ potential is depicted with extended domain and range in Fig.~\ref{fig:WKBTruncation}(a). The force potentials support the identified quantum phase transition at $p=2$, at which the central peak in the force potential bifurcates. In addition, the force potentials indicate that quantum reflection increases for values above $p=4$, reaching peak heights above one for $p=7.8$, one hundred near $p=50$, and one thousand near $p=150$, which supports the quantum-scattering finding that deeper energy dips occur at RSM energies for larger $p$ values.

The extended image of the  WKB force potential as a function of the truncation length $L$ provided in Fig.~\ref{fig:WKBTruncation}(b) further supports the use of the truncated  $V(x)=-\lvert x\rvert ^4$ potential with sufficiently long $L$ to examine scattering from the infinite-length $V(x)=-x^4$ potential. For truncation lengths beyond the main region of reflection from the infinite-length  potential (approximately $\lvert x\rvert \le3$), the largest magnitude of the force potential decreases as the truncation length $L$ increases. This finding agrees with the prediction that quantum reflection is minimized where the wavelength is short compared to the rate of change of the potential with position. In addition, the truncated and infinite-length WKB force potentials are found to agree in the domain $\lvert x\rvert <L$. Since the force potentials tend towards zero in the large-$x$ limit, truncated potentials with sufficiently long $L$ capture a large degree of the quantum reflection due to the infinite-length potential.  

\section{Robustness of RSM Predictions}\label{sec:Noise}

To demonstrate that RSMs  of the truncated $V(x)=-\lvert x\rvert ^p$ potentials are effective for experimental measurements of $\mathcal{PT}$-symmetry behavior, we demonstrate the RSMs are robust to various types of experimental errors. We show that the RSMs of the truncated $p=4$ potential reproduce the weakly-bound-state energies of the upside-down, $\mathcal{PT}$-symmetric $p=4$ potential in the presence of four types of error.

The effect of random noise that conserves $\mathcal{PT}$ symmetry is modeled as a sum of sinusoidal random Fourier coefficients of the form:
\begin{align}
    V_\text{R}(x)&=-(x^{4}+n_\text{R}g(x))f\left(x,w,L\right)-L^{4}f\left(-x,w,L\right),\nonumber\\
    g(x)&=\sum^N_{i=1} a_i \cos\left(\frac{2\pi }{\omega_i}x\right),\label{eq:SymRandomNoise}
\end{align}
where $f$ is the smoothing function (see Methods), $N=50$ is the number of random Fourier components, $a_i\in[-1,1]$ is a random amplitude, $\omega_i\in[0.1,5]$ is a random frequency, and $n_\text{R}$ is the strength of the random noise. Parity-breaking random noise is likewise modeled with sinusoidal random Fourier-coefficient noise of the form
\begin{align}
g(x)&=\sum^N_{i=1} a_i \sin\left(\frac{2\pi }{\omega_i}x+\phi_i\right)\label{eq:RandomNoise}
\end{align}
where $\omega_i\in[0.1,5]$ is a random frequency. 
Experimental errors that result in polynomial variation of the potential are modeled with an additive negative quadratic potential term
\begin{align}
    V_\text{NQ}(x)&=-(x^{4}+n_\text{NQ}x^2)f\left(x,w,L\right)\nonumber\\
    &-(L^{4}+n_\text{NQ}L^2)f\left(-x,w,L\right),\label{eq:NQNoise}
\end{align}
or a positive quadratic term
\begin{align}
    V_\text{PQ}(x)&=-(x^{4}-n_\text{PQ}x^2)f\left(x,w,L\right)\nonumber\\
    &-(L^{4}-n_\text{PQ}L^2)f\left(-x,w,L\right),\label{eq:PQNoise}
\end{align}
where $n_\text{NQ}$ and $n_\text{PQ}$ are the the negative and positive quadratic term strengths.

\begin{figure*}
\centering
\includegraphics[height=0.27\textheight]{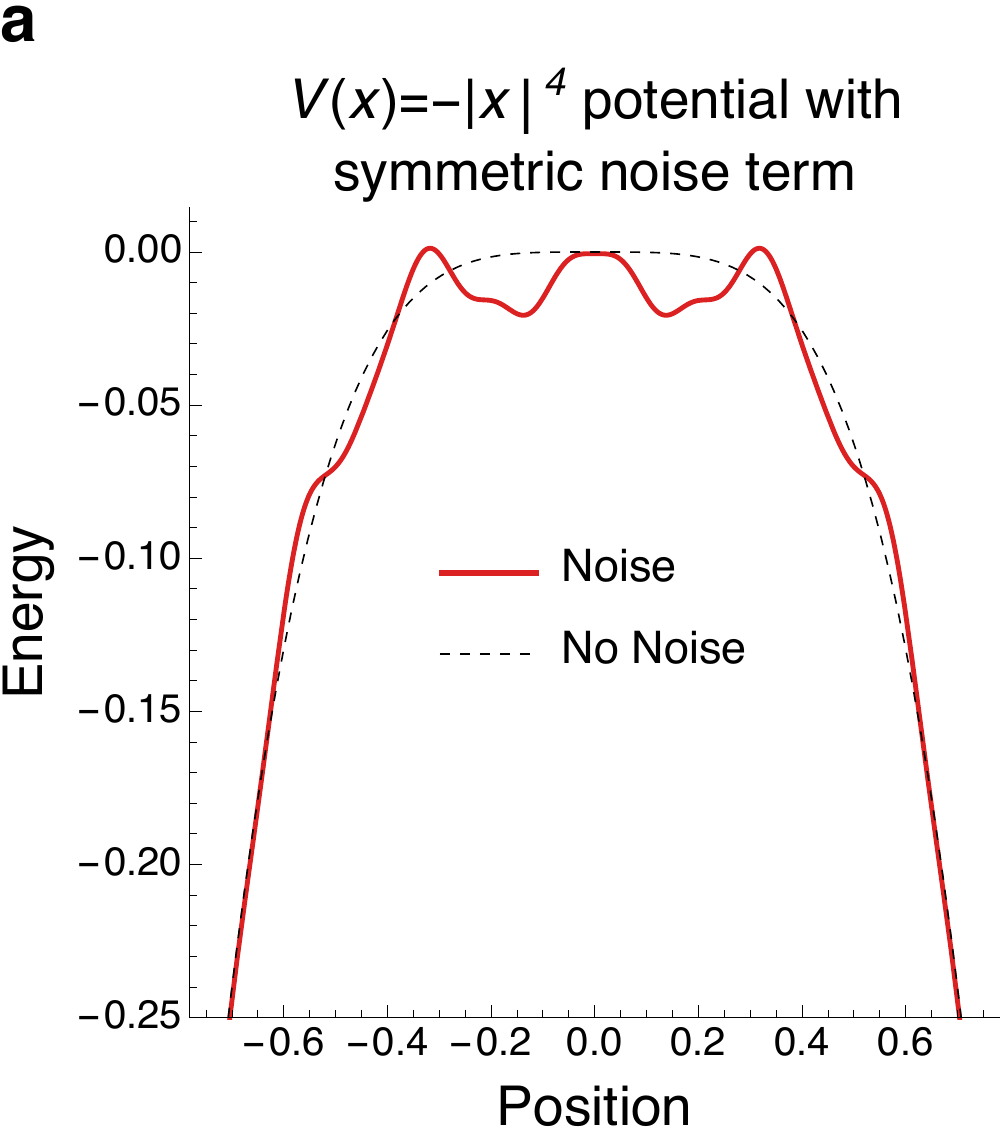}\,\,\,\,\,\includegraphics[height=0.25
\textheight]{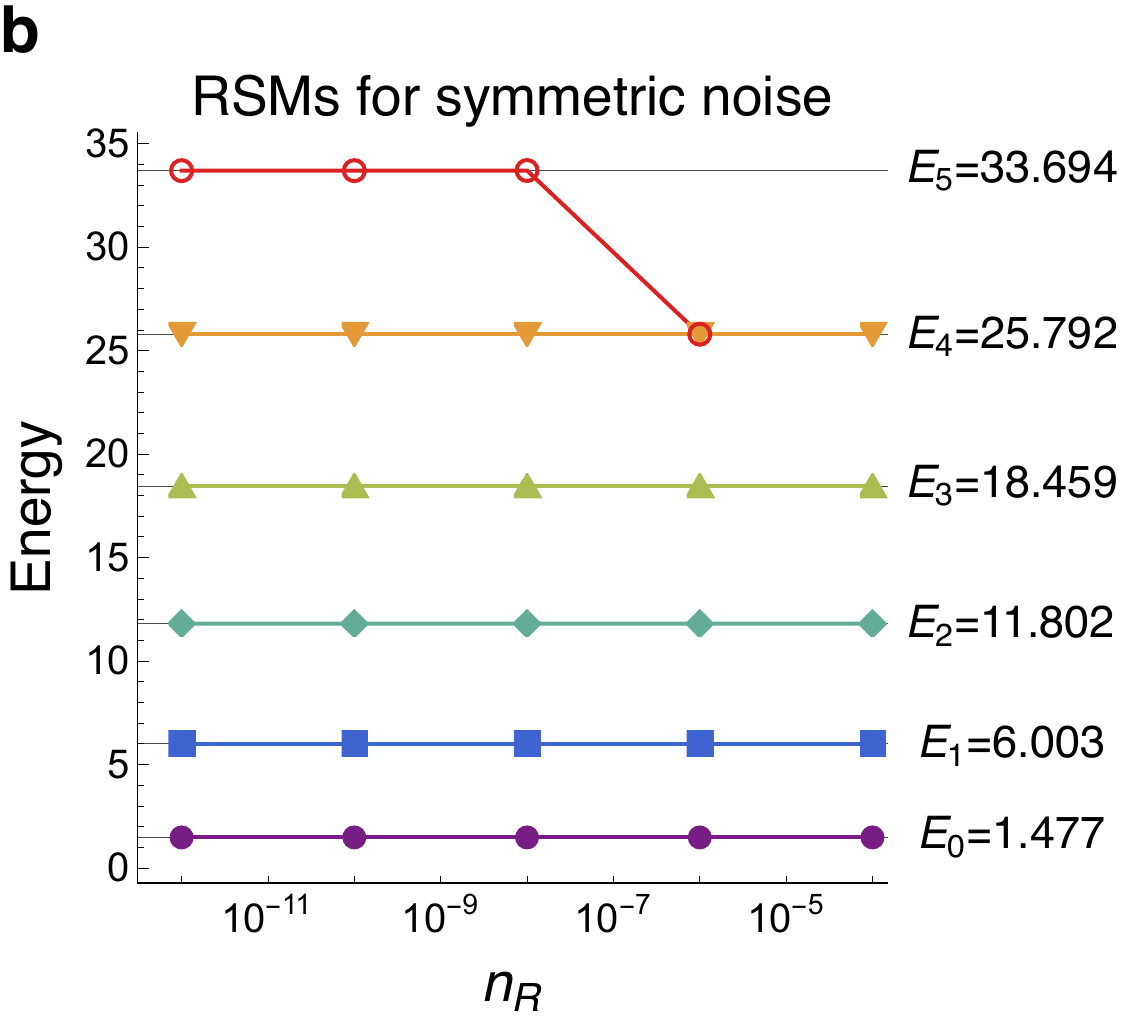}
\caption{(a) Close-up near the origin of a truncated $V(x)=-\lvert x\rvert ^4$ potential (dashed black line) with symmetric random noise Eq.~\eqref{eq:SymRandomNoise} of strength $n_\text{R}=0.01$ (solid red line) and its (b) RSMs as a function of noise strength  $n_\text{R}$. The RSMs for the noisy potential  (colored lines with points) are found to closely agree with those without noise (horizontal gray lines, $E_i$ for $i=0-5$) over several orders of magnitude of $n_R$. The RSMs are averaged for three random sets of parameters $a_i$ and $\omega_i$.
}\label{fig:SymmetricNoise}
\end{figure*}

Figure~\ref{fig:SymmetricNoise} indicates that the RSMs of the model with symmetric random noise closely agree with the known weakly-bound states. The lowest-lying eigenenergies agree with the RSMs for all noise strength terms considered, and higher eigenenergies are reproduced by further decreasing the noise strength.

As shown in Fig.~\ref{fig:SymmetricNoise}, the R-zeros accurately reproduce the weakly-bound-state energies even in the presence of parity-breaking random noise. As expected for R-zeros in time-reversal-symmetric systems, the R-zeros lie off the real axis. The real parts of the R-zeros closely agree with the known RSMs of the exact $\mathcal{PT}$-symmetric potential. Even when random noise causes significant variation in the potential near the origin, 
the low-lying eigenenergies are accurately recovered for noise strengths that vary over orders of magnitude. Higher energies are recovered as the noise strengths are decreased. In addition, the distance from the real axis scales according to a power law with the parity-breaking noise strength, such that symmetry breaking does not significantly affect the RSMs for sufficiently small errors.

Likewise, additive quadratic terms that significantly change the form of the potential yield near-reflectionless scattering modes that closely agree with the low-lying exact weakly-bound-state energies for a wide range of quadratic term strengths (see Fig.~\ref{fig:QuadraticNoise}). And, higher-lying energies are determined accurately for smaller quadratic term strengths.

\begin{figure*}
\centering
\includegraphics[height=0.25\textheight]{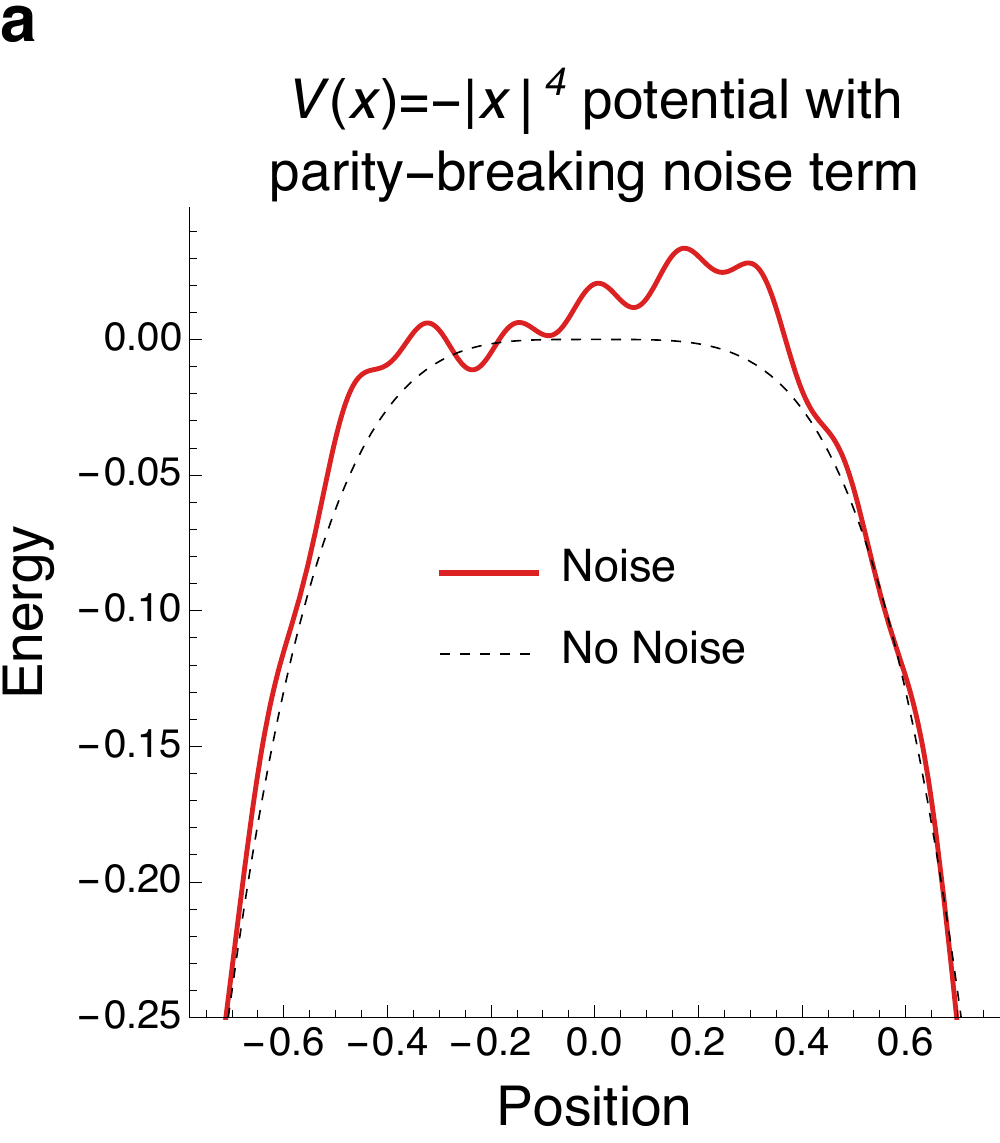}\,\,\includegraphics[height=0.22
\textheight]{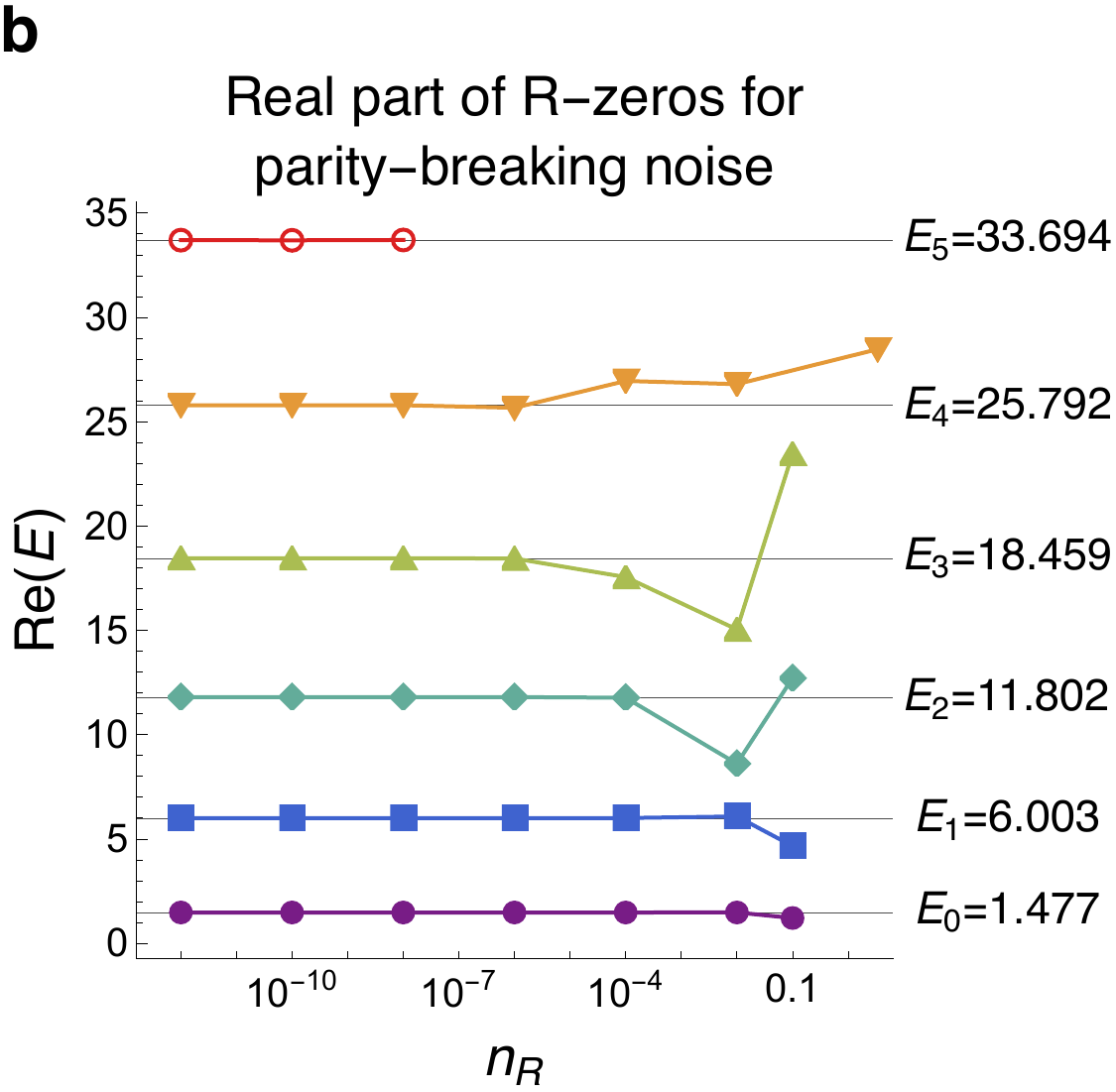}\,\,\,\,\,\,\includegraphics[height=0.22
\textheight]{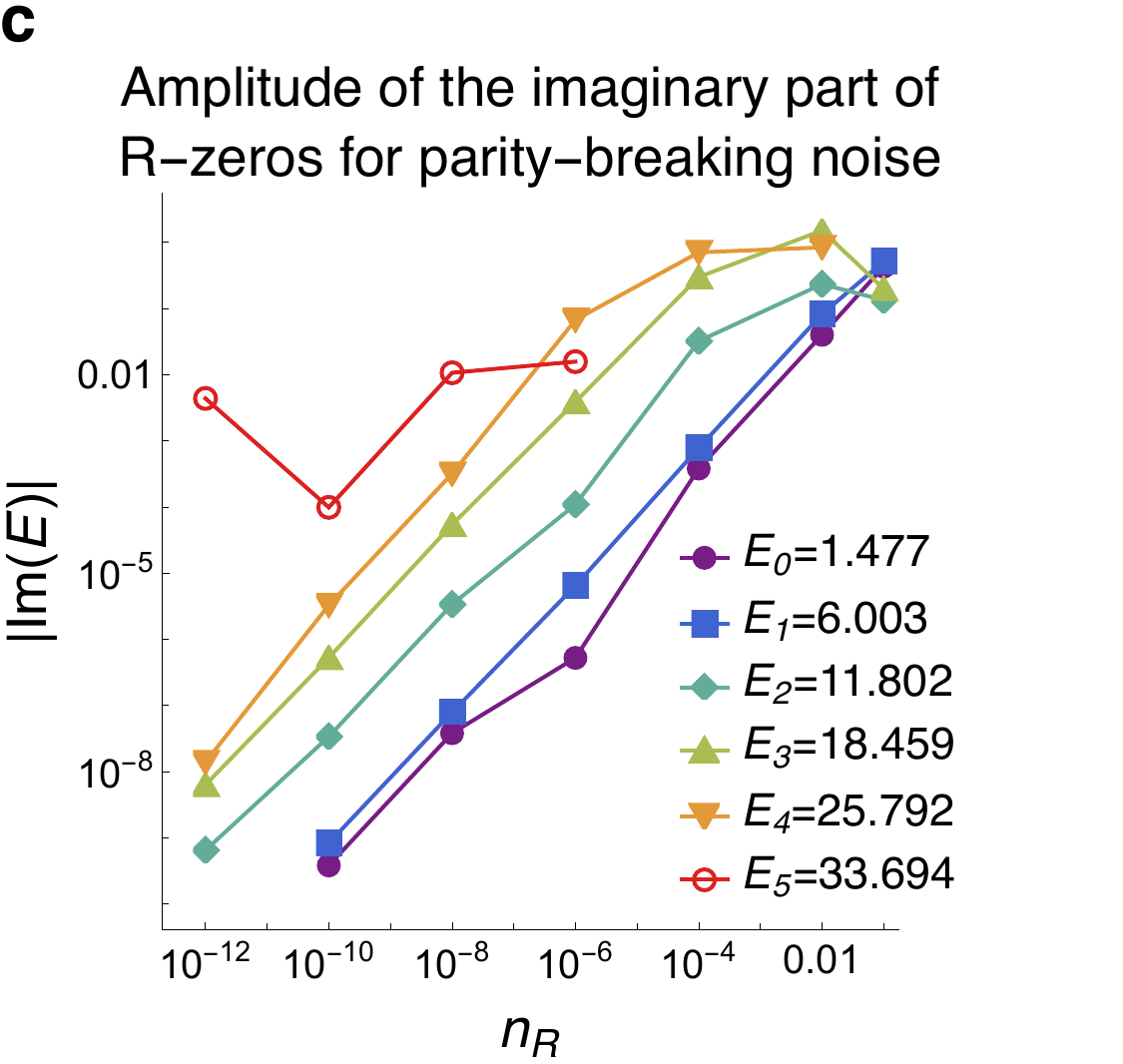}
\caption{(a) Detail of truncated $V(x)=-\lvert x\rvert ^4$ potential (dashed black line) with parity-breaking random noise Eq.~\eqref{eq:RandomNoise} (solid red line) near the origin (noise strength $n_\text{R}=0.01$ pictured) and the (b) real part and (c) amplitude of the imaginary part of its R-zeros (with imaginary parts lower than threshold value $\varepsilon=3$) for varying  $n_\text{R}$. The real part of the R-zero in the presence of the noise (colored lines with points) closely matches the noise-free result (horizontal gray lines, $E_i$ for $i=0-5$) over several orders of magnitude in strength term $n_R$, and the imaginary part of the R-zero (colored lines with points to corresponding $E_i$ for $i=0-5$) due to parity breaking approaches zero as $n_R$ is decreased. Results are averaged over three random sets of parameters $a_i$, $\omega_i$, and $\phi_i$.
}\label{fig:GaussianNoise}
\end{figure*}

\begin{figure*}
\centering
\includegraphics[height=0.25
\textheight]{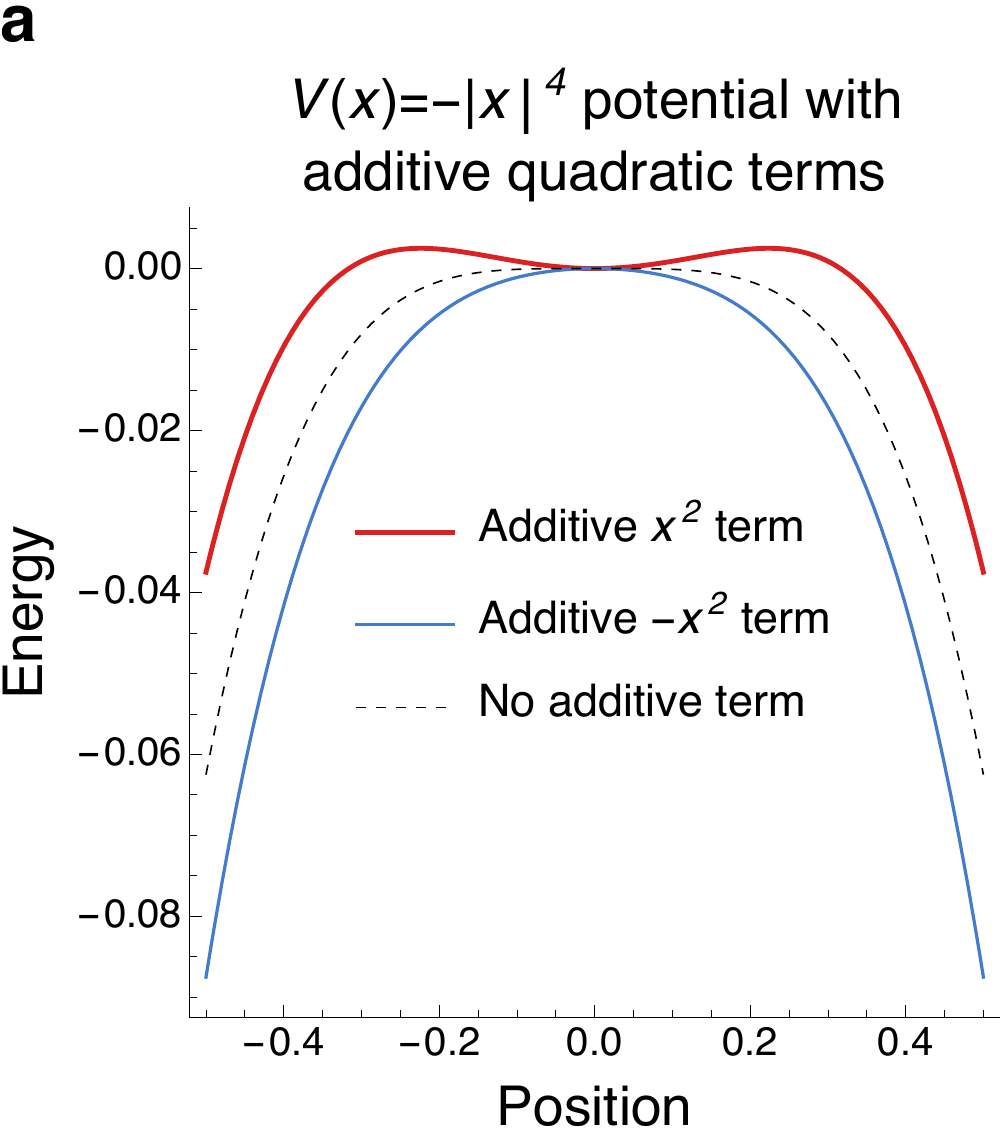}\,\includegraphics[height=0.22\textheight]{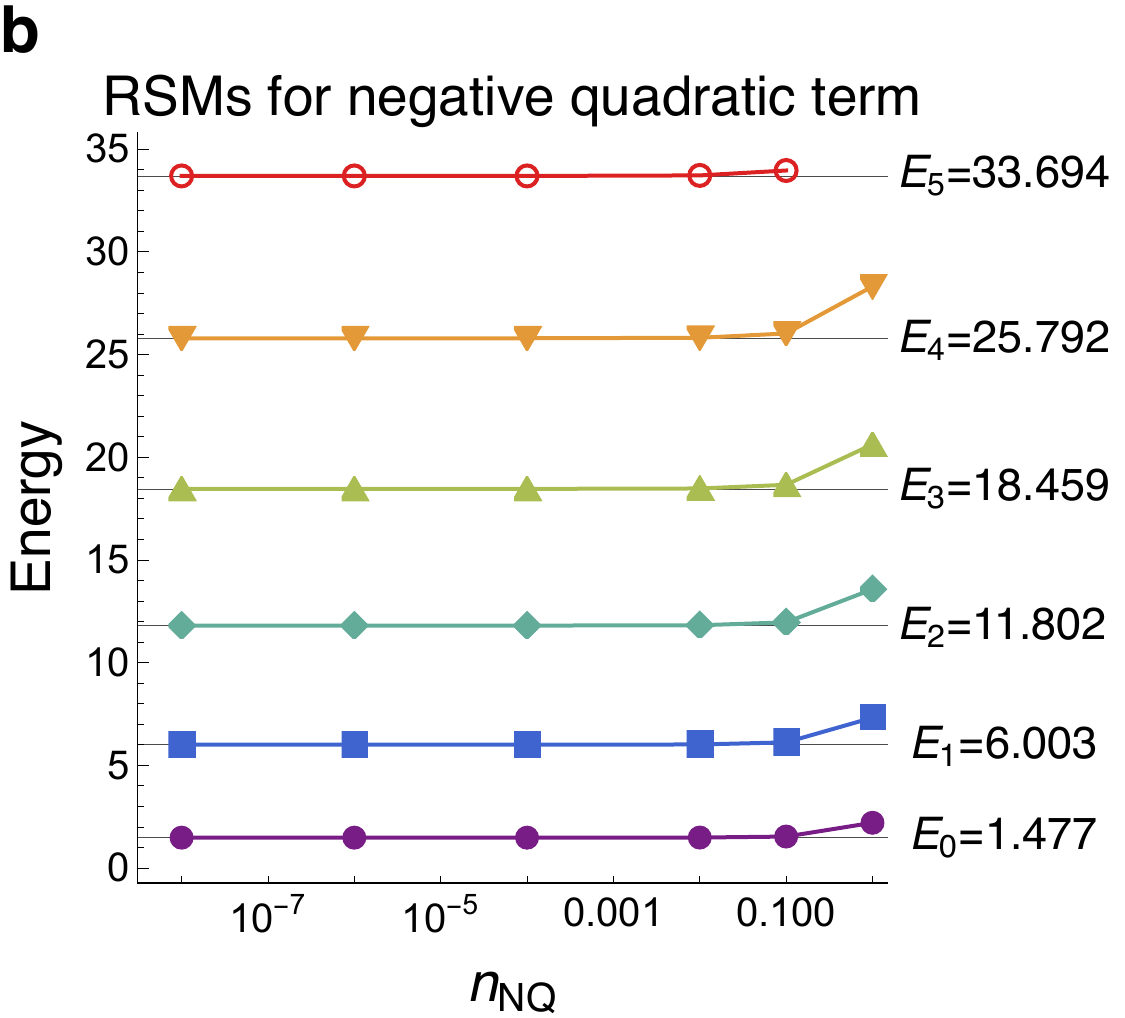}\includegraphics[height=0.22\textheight]{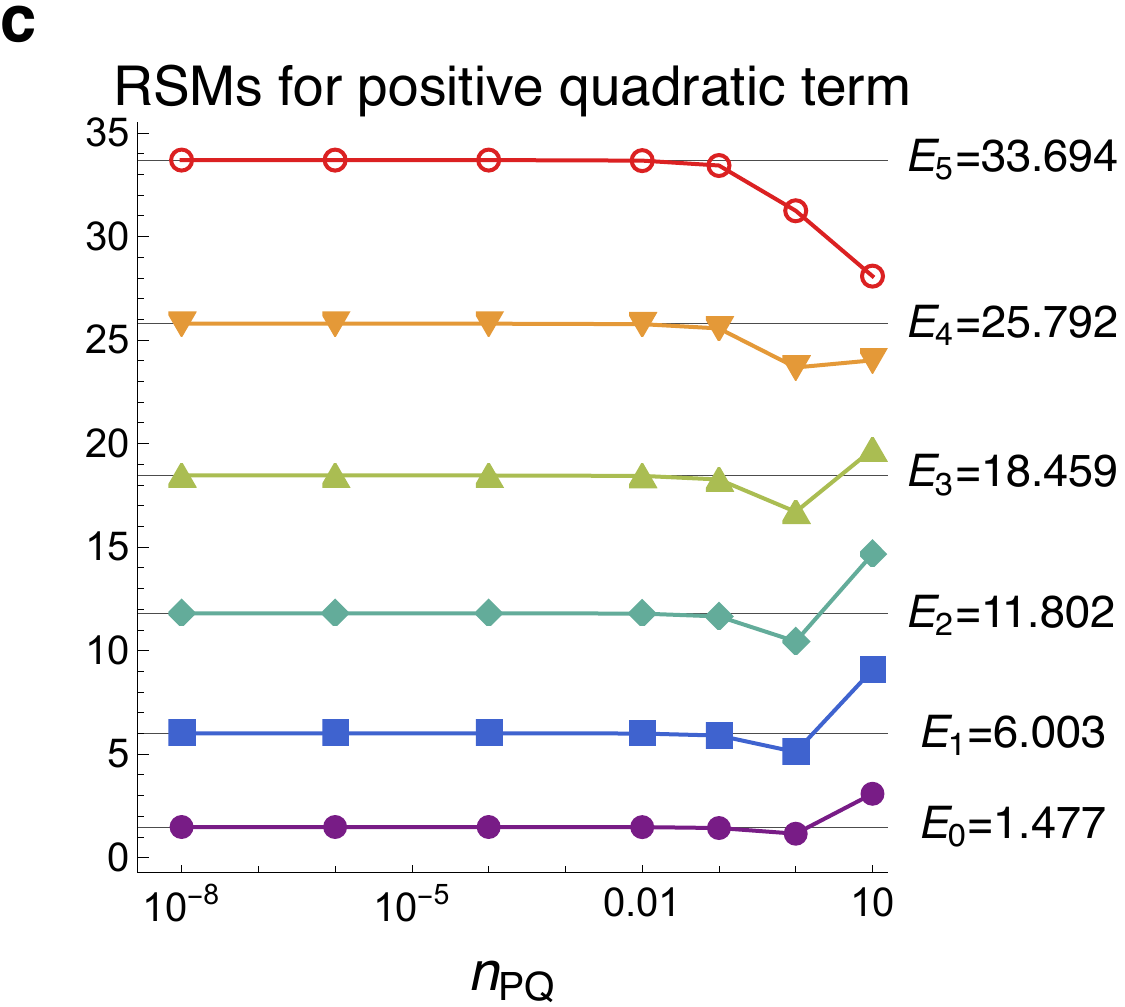}
\caption{(a) Detail of truncated $V(x)=-\lvert x\rvert ^4$ potential near the origin with additive quadratic terms Eq.~\eqref{eq:NQNoise} and Eq.~\eqref{eq:PQNoise}. Potentials are shown for a positive quadratic term with strength $n_\text{PQ}=0.1$ (thick solid red line), negative quadratic term with strength $n_\text{NQ}=0.1$ (medium solid blue line), and no additive term (thin dashed black line).  The RSMs in the presence of additive quadratic terms (colored lines with points) reproduce those of the potential with no additive term (horizontal gray lines for $E_i$ with $i=0-5$) over a broad range of negative $n_{NQ}$ and positive $n_{PQ}$ quadratic term strengths. RSMs are shown as a function of the term strength (b) $n_\text{NQ}$ and (c) $n_\text{PQ}$. \label{fig:QuadraticNoise}}
\end{figure*}

\end{document}